\documentclass[%
reprint,
superscriptaddress,
aps,
]{revtex4-1}

\usepackage{graphicx}
\usepackage{hyperref}
\hypersetup{colorlinks=true, linkcolor=dgreen, urlcolor=dblue,
citecolor=dgreen}
\usepackage{amsmath} \usepackage{amssymb}
\usepackage{accents} \usepackage{mathrsfs} \usepackage{setspace}
\usepackage{color}
\definecolor{dgreen}{RGB}{00, 120, 00} \definecolor{dblue}{RGB}{00, 00, 180}
\definecolor{lgreen}{RGB}{46, 139, 87} 
\usepackage{ulem}

\newcommand{\bs}[1]{\boldsymbol{#1}}

\newcommand{\ut}[1]{\undertilde{#1}}

\begin{document}

%
%

\title{Vortex supercurrent inversion by frequency-symmetry conversion
of Cooper pairs}

\author{Soma Yoshida}
\affiliation{Department of Applied Physics, Nagoya University, Nagoya 464-8603, Japan }
\author{Shu-Ichiro Suzuki}
\email[Corresponding author: ]{s.suzuki-1@utwente.nl}
\affiliation{MESA+ Institute for Nanotechnology, University of Twente, 7500 AE Enschede, The Netherlands}
\affiliation{Department of Applied Physics, Nagoya University, Nagoya 464-8603, Japan }
\author{Yukio Tanaka}
\affiliation{Department of Applied Physics, Nagoya University, Nagoya 464-8603, Japan }

\date{\today}

\begin{abstract}
  We theoretically demonstrate that the vortex supercurrent can be
	reversed by 
	odd-frequency Cooper pairs accompanied by surface
	Andreev bound states. The surface of a three-dimensional superconductor
	pierced by a flux quantum is considered. 
	We compare the vortex supercurrents near the surface of the
	spin-singlet $s$-wave and spin-triplet $p_z$-wave superconductors
	using quasiclassical Eilenberger theory, 
	where the surface is perpendicular to the $z$ direction.
	We demonstrate that the
	vortex supercurrent near the surface of a $p_z$-wave superconductor
	is reversed compared to those far from the surface, whereas that of
	an $s$-wave superconductor is not. 
	The splitting of the zero-energy states caused by the 
	interference of the surface Andreev bound states and 
	Caroli-de Gennes-Matricon modes is demonstrated. 
\end{abstract}

\pacs{pacs}

\thispagestyle{empty}

\maketitle

\section{Introduction}

The magnetic flux penetrating a type-II superconductor (SC) is quantized
(i.e., quantum vortices) since a macroscopic wave function describing
the Cooper pair condensate must be single-valued.  The phase winding
around the vortex results in a shielding current that localizes the
magnetic field at the core. This is how type-II SCs maintain the
perfect diamagnetism even in a magnetic field. 

However, recent experiments confirmed the existence of
\textit{paramagnetic} superconducting states\cite{Bernardo_PRX_2015,
Krieger_PRL_2020}. Applying a low-energy muon spin relaxation
($\mu$SR) technique for superconducting bilayer systems, the local
magnetic field was larger than that of the applied magnetic field.
This local enhancement is well-understood in terms of odd-frequency
(odd-$\omega$) Cooper pairs \cite{Berezinskii, balatsky_1992_PRB,
Linder_2019_RMP}. The pair density of odd-$\omega$ pairs was
\textit{effectively negative}\cite{Asano_PRL_2011}, meaning that
odd-$\omega$ pairs show the opposite response to perturbations
compared with the well-known even-$\omega$
pairs\cite{Bernardo_PRX_2015, Krieger_PRL_2020, Asano_PRL_2011,
Yokoyama_PRL_2011, Higashitani_PRL_2013, Higashitani_PRB_2014,
Suzuki_PRB_2014, Suzuki_PRB_2015, Asano_PRB_2015}: Novel odd-$\omega$
pairings attract external fields. 

Odd-$\omega$ pairs appear in inhomogeneous
systems\cite{tanaka_PRL_1995, tanaka_PRB_2005, tanaka_PRL_2007,
Eschrig_Nat_2007, Suzuki_PRB_2016, Asano_PRL_2007, tanaka_JPSJ_2012,
Suzuki_PRR_2021, rogers2021spin}.  In addition to $U(1)$-symmetry
breaking, the inhomogeneity locally breaks the inversion symmetry and
causes parity mixing of the Cooper pairs. Parity-converted Cooper
pairs should belong to an odd-$\omega$ symmetry class to satisfy the
Fermi-Dirac statistics (i.e., Berezinskii rule \cite{Berezinskii}). 
For example, in a vortex core in a BCS-type SC, odd-$\omega$ pairs are
induced because of the vortex singularity\cite{yokoyama_PRB_2008,
tanuma_PRL_2009}.  At the surface of a spin-triplet $p$-wave SC,
odd-$\omega$ pairs appear \cite{tanaka_PRB_2005, tanaka_PRB_2007,
eschrig_JLTP_2007}, where surface Andreev bound
states\cite{hara_PTP_1986} (ABSs) strongly suppress the local pair
potential. 

\begin{figure}[b]
	\centering
	\includegraphics[width=0.9\columnwidth]{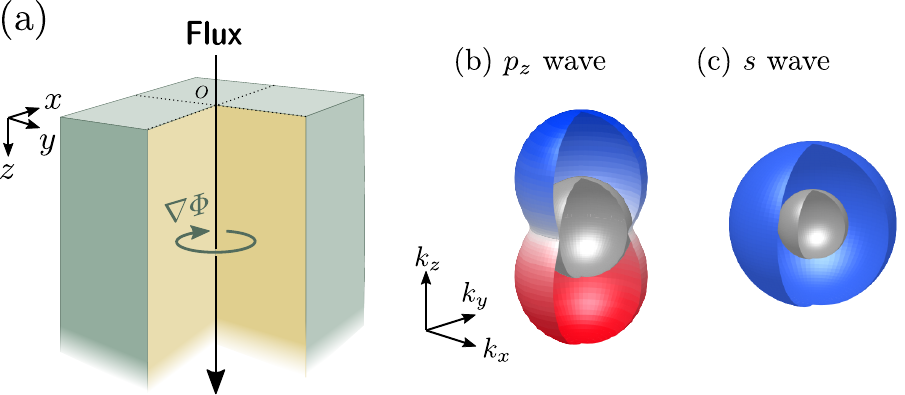}
	\caption{ (a) Schematic of the system. The vortex core is
	located where $x=y=0$ and the semi-infinite 3D SC that occupies $z \geq 0$. 
	The gap amplitudes for the (b) $s$- and (c) $p_z$-wave SC, where
	color means the internal phase of the pair potential. The inner
	silver sphere indicates the Fermi sphere.} 
  \label{model}
\end{figure}

Odd-$\omega$ pairs in a vortex core are generated by a singularity in
real space, whereas odd-$\omega$ pairs at the surface generate 
ABSs, which are related to anisotropy in the momentum space. 
The two types of
odd-$\omega$ pairs were considered
separately\footnote{The vortex shadow effect\cite{
graser_PRL_2004, yokoyama_PRL_2008, silaev_PRB_2009} was discussed. However,
odd-$\omega$ pairs are not in the same location. }. However, these
odd-$\omega$ pairs can emerge at the same location and interact with each other.
Recently, $k_z$-dependent anisotropic pairings in
three-dimensional (3D) SCs have become realistic (e.g., heavy-fermion
compounds\cite{ran_science_2019} and
Sr$_2$RuO$_4$\cite{pustogow_Nature_2019, Agterberg_PRR_2020,
Grinenko_NatPhys_2021, grinenko2021unsplit}). In these systems, a
quantum vortex can penetrate a two-dimensional (2D) surface hosting
flat-band ABSs \cite{Shingo_PRB_2015,TamuShun_PRB_2017,suzuki_PRB_2020,
suzuki2022}. It is unclear how odd-$\omega$ pairs by ABSs affect the
odd-$\omega$ pairs carrying the vortex supercurrent, and vice
versa. Such novel vortex states help determine the pairing
symmetry of an SC and can be easily examined by existing experimental
methods such as scanning tunneling microscopy\cite{hess_PRL_1989} (STM) and 
superconducting quantum interference devices\cite{squid1, squid2} (SQUID).

In this study, the interference between odd-$\omega$ pairs with two
different origins is investigated: the quantum vortex and surface
ABSs.  A quantum flux penetrating a semi-infinite 3D unconventional SC
is considered, where the magnetic flux is perpendicular to the
surface, at $z=0$. 
Using the quasiclassical Eilenberger theory, the vortex supercurrents
and local density of states (LDOS) in the $s$- and $p_z$-wave SCs are
compared, where surface ABSs are, respectively, absent and present.

The numerical results show that the vortex supercurrent was reversed
near the surface where odd-$\omega$ pairs are dominant.
The pair amplitude is analyzed and it is shown that there is local symmetry
conversion among the Cooper pairs in the region where the vortex and the
ABSs coexist. 
The LDOS, specifically, how the Caroli-de\,Gennes-Matricon (CdGM)
state\cite{caroli_PL_1964, hess_PRL_1989, gygi_PRB_1991} and surface
ABSs develop when they overlap is investigated. It is also shown that
the zero-energy peak in the LDOS split because of the interference. 

In the analytical calculations, the knowledge gained from the numerical
results is extended to more general SCs.
By using the Kramer-Pesch approximation\cite{kramer_ZP_1974,
nagai_JPSJ_2006, nagai_JPSJ_2019}, which is valid at a low
temperature and near the vortex core, the direction of the
vortex supercurrent near the surface is shown to be determined only by
the $k_z$ dependence of the pair potential.

The remainder of this paper is organized as follows. The model and method are
introduced in Sec.~\ref{modelsec}. The numerical and analytical 
results are presented in Sec.~\ref{resultnumsec} and Sec. ~\ref{resultanasec}. 
In Sec.~\ref{conclusionsec}, the study is summarized.

\section{System and formulation}\label{modelsec}

We considered the quantum vortex penetrating a 3D
SC as shown in Fig.~\ref{model}(a). The SC occupies $z \geq 0$ and the
vortex core is located at $x=y=0$. We assume the cylindrical symmetry
around the core. We employ the
cylindrical coordinate: $(x, y, z) = (\rho \cos \Phi, \rho \sin \Phi,
z)$.

The superconductivity in the ballistic limit can be described by
quasiclassical Eilenberger theory\cite{eilenberger_ZP_1968,
eilenberger_ZP_1969, larkin1969quasiclassical,
schopohl_PRB_1995,schopohl1998transformation} . The
quasiclassical Green's function obeys the Eilenberger equation:
\begin{align}
  &\hbar \bs{v}_F \cdot \bs{\nabla} \hat{g}
	+[\hat{H},\hat{g}]=0,
	\quad
	\hat{g}^2= \hat{1} \label{eilenberger}
	\\
  &\hat{g}=\left(
    \begin{array}{cc}
      g & -s_\mu f \\
      -\ut{f} & -\ut{g}
      \end{array}
  \right),
  \quad
  s_\mu=\biggl\{
	  \begin{array}{ll}
			+1 & \text{for $s$ wave, }\\
			-1 & \text{for $p_z$ wave, }
	  \end{array}\nonumber\\
	  &\hat{H}=\left(
    \begin{array}{cc}
			\omega_n-i(e/c) \bs{v}_F\cdot\bs{A} & 
			-s_\mu\Delta \\
	    - \ut{\Delta} & 
			-\omega_n+i(e/c)\bs{v}_F\cdot\bs{A}
      \end{array}
  \right),\nonumber
\end{align}
where 
$\bs{v}_F = v_F \hat{\bs{k}}$ with
$\hat{\bs{k}} = (k_x, k_y, k_z) = 
(
\sin \theta \cos \phi,  
\sin \theta \sin \phi, 
\cos \theta
)$, 
$\omega_n=(2n+1)k_BT$, $T$, $e<0$, $c$,
and $\bs{A}$ are the isotropic Fermi velocity, 
Matsubara frequency, 
charge of a quasiparticle, speed of light, and vector potential. 
The normal
and anomalous Green's functions 
[$g(\hat{\bs{k}},\bs{r},i\omega_n)$ and $f(\hat{\bs{k}},\bs{r},i\omega_n)$]
describe the
quasiparticle and the Cooper pairs. 
The under-tilde functions have been introduced as 
$\ut{X}(\hat{\bs{k}},\bs{r},i\omega_n)=X^*(-\hat{\bs{k}},\bs{r},i\omega_n)$
with X being an arbitral function. 
In this paper, we refer to $f$ as the pair amplitude. 
In Eq.~\eqref{eilenberger}, we have reduced the spin degree of freedom
by assuming the opposite-spin Cooper pairing for both SCs.

\begin{figure*}[tbp]
	\includegraphics[width=1.9\columnwidth]{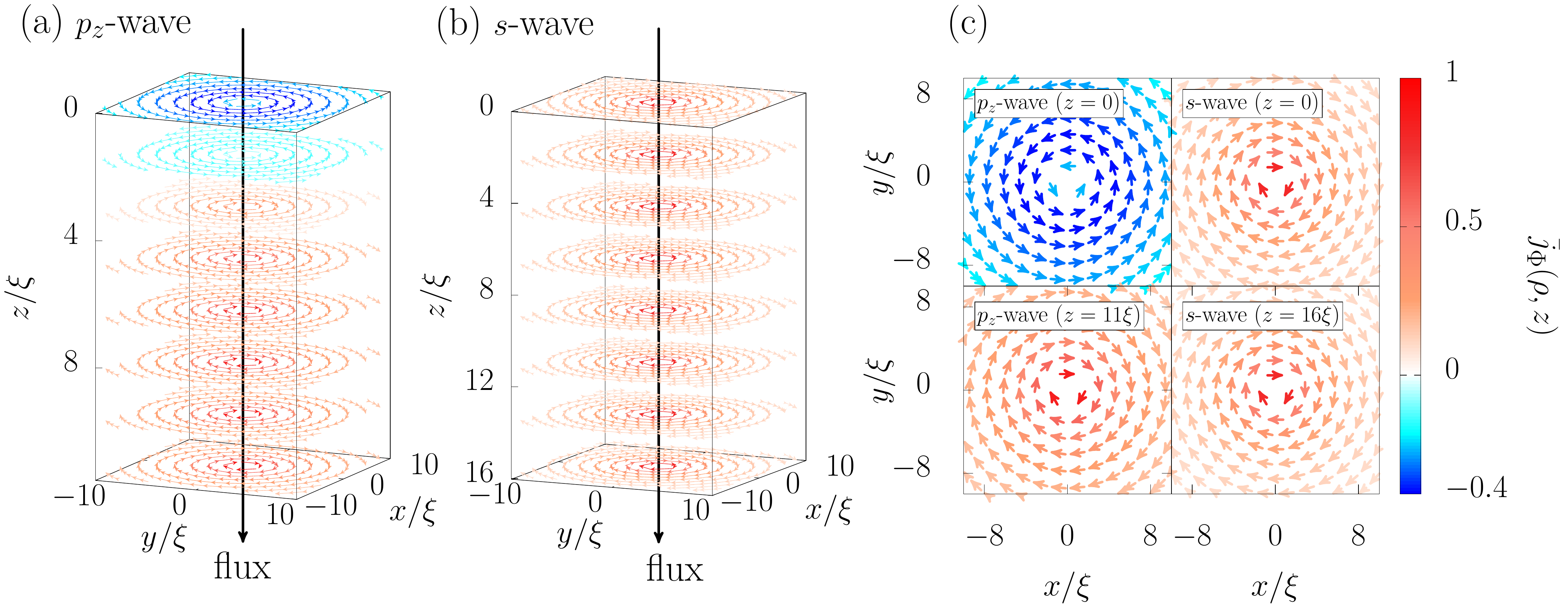}
	\caption{Vector plots of the current densities at $T/T_c=0.2$ for
		(a) the $p_z$-wave and (b) $s$-wave SCs. The direction and color
		of the arrows mean the direction and
		azimuthal component of the current density. The current density is
		normalised as $\bar{j}_\Phi =
		j_\Phi/j^{\mathrm{Max}}_\Phi(T/T_c=0.2)$ with
		$j^{\mathrm{Max}}_\Phi(T/T_c=0.2)$ being 
		the maximum value of $j_\Phi$ at
		$T/T_c=0.2$ in each case. 
		The solid black arrows represent the magnetic flux. 
		(c) Top view of the same
		results at $z=0$ and $z\gg\xi$.  
		The parameters are set to $\Omega_{c}=40k_BT_c$ and 
		$\kappa_{\mathrm{GL}}=6\sqrt{6}$.}
	\label{fig:cur}
\end{figure*}

We consider the spin-singlet $s$-wave and spin-triplet $p_z$-wave SCs
whose pair potentials can be given by 
\begin{align}
	\Delta(\bs{r}, \hat{\bs{k}})
	= \left\{ \begin{array}{ll}
			\Delta(\bs{r})    & \text{for $s$ wave,} \\
		  \Delta(\bs{r})k_z & \text{for $p_z$ wave.}
	\end{array} \right.
\end{align} 
The schematic gap functions are shown in Fig.~\ref{model}. 
The pair potential $\Delta(\bs{r})$ is self-consistently determined by the 
gap equation: 
\begin{align}
	\Delta(\hat{\bs{k}},\bs{r})
	&=2\lambda N_0 \pi k_B T\sum_{n=0}^{N_{\mathrm{c}}}
	\int\frac{d\Omega^\prime}{4\pi}
	V(\hat{\bs{k}},\hat{\bs{k}}^\prime)
	f(\bs{k}^\prime,\bs{r},i\omega_n),
	\label{gapeq}
	\\
	\lambda
	&=N_0^{-1} \left(\sum_{n=0}^{N_{\mathrm{c}}}
	\frac{1}{n+1/2}+\ln\frac{T}{T_c}\right)^{-1},\nonumber
\end{align}
where $\int \cdots d\Omega/4\pi$ is the average on the Fermi sphere, 
$N_0$ is a density of states (DOS) at Fermi energy in the normal
state, and $N_c$ is defined as the positive integer satisfying
$N_c<\Omega_{c}/2\pi k_BT$ with 
$\Omega_{c}$ being the cut-off
frequency. 
The attractive potential $V$ depends on the pairing symmetry:	
$V=1$ for the $s$-wave SC and 
$V=3 k_z k_z'$ for the $p_z$ wave SC. 
The pair potential winds its phase around the vortex; $\Delta(\bs{r})
= \Delta(z, \rho) e^{-i \Phi}$ with 
$(x,y,z)=(\rho\cos\Phi,\rho\sin\Phi,z)$. 

The current density $\bs{j}(\bs{r})$ and LDOS $N(\bs{r},\omega)$ are given by
\begin{align}
	&\bs{j}(\bs{r})
	=-4|e|N_0\pi k_BT\sum_{n=0}^{N_{\mathrm{c}}}
	\int\frac{d\Omega}{4\pi}
	\bs{v}_F \cdot \mathrm{Im}g(\hat{\bs{k}},\bs{r},\omega_n), 
	\label{jeq}
	\\
	& \frac{N(\bs{r},\omega)}{N_0}
	=\lim_{\delta\rightarrow+0}
	\int\frac{d\Omega}{4\pi}\mathrm{Re} 
	\left[ g(\hat{\bs{k}},\bs{r},\omega+i\delta) \right]. 
\end{align}
where $\delta$ is the smearing factor.
The vector potential is also self-consistently determined
using the following equations:
\begin{align}
	& \bs{\nabla} \times \bs{A} = \bs{B}, 
	\hspace{6mm} 
	\bs{\nabla} \times \bs{B} = \frac{4\pi}{c}\bs{j}.\label{maxeq}
\end{align}
In the quasiclassical theory,  
the Ginzburg-Landau (GL) parameter $\kappa=\lambda_L/\xi$ is a
parameter that
characterizes the length scale of the magnetic field, 
where $\lambda_L=\sqrt{3c^2/8\pi e^2v_F^2N_0}$ and $\xi=\hbar v_F/2\pi k_B T_c$ are 
the London penetration depth and coherence length (see the Appendix \ref{sec:form} for the detail). 
In the numerical calculations, we 
set $\kappa=6\sqrt{6}$ and $\Omega_c=40k_B T_c$.

\section{Numerical Results}\label{resultnumsec}

\subsection{Paramagnetic quantum vortex}
\begin{figure}[tbp]
	\centering
	\includegraphics[width=0.98\columnwidth]{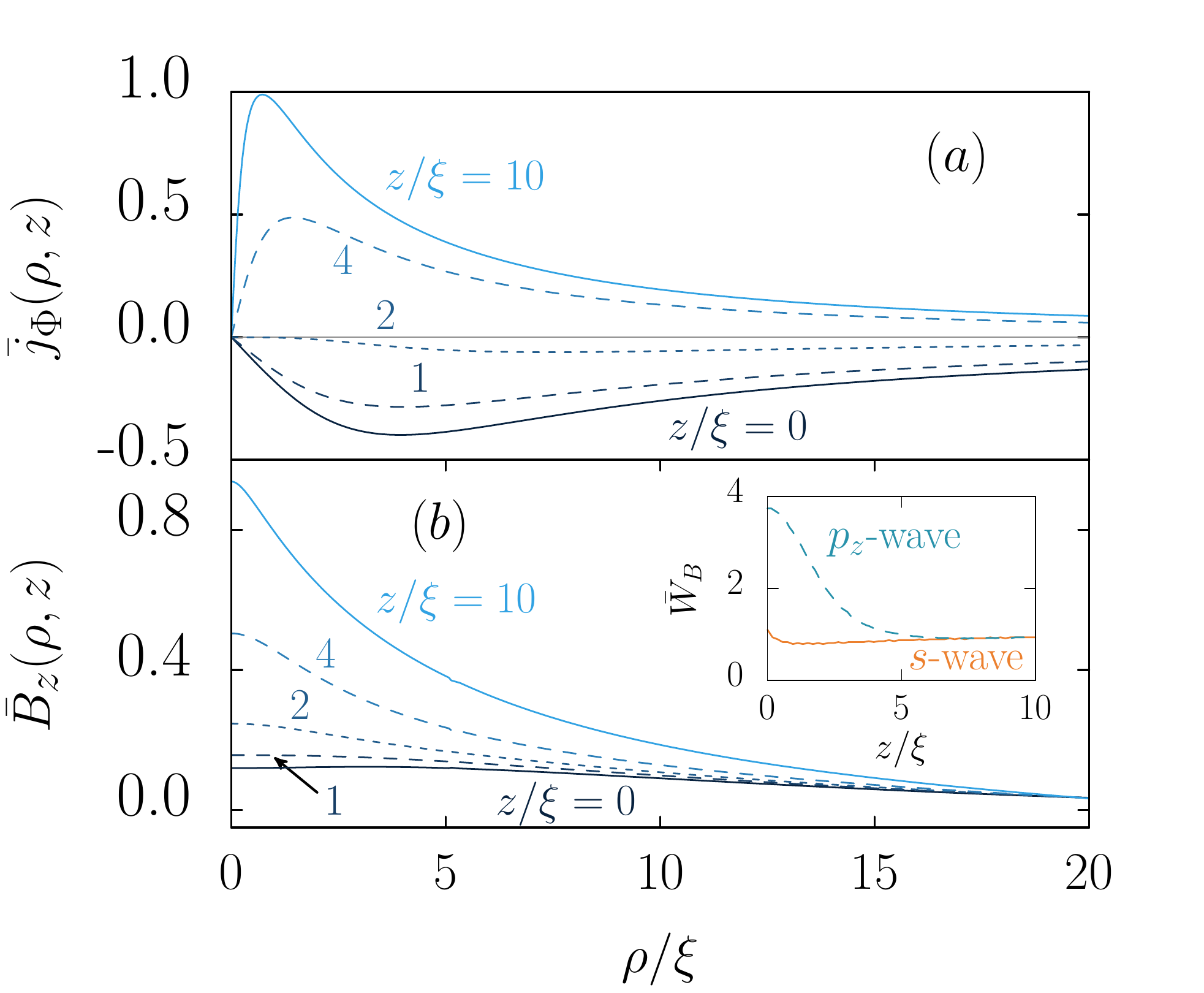}
	\caption{Spatial profiles of (a) the circulation current and (b) the
		magnetic field in the $p_z$-wave SC. The magnetic field is
		normalized as $\bar{B}_z = B_z/B_z(\rho=0,z=10\xi,T/T_c=0.2)$. The depth is set to $z/\xi =
		0, 1, 2, 4$, and $10$. Near the surface, the current changes the direction 
		and the magnetic field is significantly suppressed. The depth
		dependence of $\bar{W}_B$ is shown in the inset.}
		\label{jprho}
\end{figure}

The vector plot of the current density around the vortex core in the
$p_z$-wave SC is shown in Fig.~\ref{fig:cur}(a), where the result for
the $s$-wave SC is shown in Fig.~\ref{fig:cur}(b) as a reference. The
color of the vectors corresponds to $j_\Phi$: the red and blue mean
the diamagnetic and paramagnetic current respectively.  At the surface
of the $p_z$-wave SC, the current density changes its direction near
the surface as shown in Fig.~\ref{fig:cur}(a), whereas $\bs{j}$ in the
$s$-wave case hardly depends on $z$ as shown in Fig.~\ref{fig:cur}(b).
The vortex supercurrent inversion can be seen more clearly in the top views
in Fig.~\ref{fig:cur}(c). 
The $\rho$ dependence of $j_\Phi$ for the $p_z$-wave SC is shown in
Fig.~\ref{jprho}(a), where we fixed the depth at $z/\xi=0, 1, 2, 4,
10$. We see that the profile of $j_\Phi$ at $z=10\xi$ is qualitatively
the save as the well-known $s$-wave result \cite{gygi_PRB_1991}.
However, with approaching to the surface, $j_\Phi$
changes the sign at $z \sim 2 \xi$, meaning that the paramagnetic
current flows only near the surface. 
The self-consistent pair potentials are shown in the
Appendix~\ref{sec:num_res}. 

The $z$ component of the local magnetic field $B_z$ for the $p_z$-wave SC 
is shown in Fig.~\ref{jprho}(b). Deep inside the SC, 
$B_z$ has the
typical profile: the magnetic field has a peak at the
vortex core and decays in the order of $\lambda_L$.
However, with approaching to the
surface, $B_z$ is strongly suppressed and 
spreads broader. Although the paramagnetic current
flows near the surface, it does not change $\mathrm{sgn}[B_z]$ but 
modifies significantly the spatial profile of $B_z$ 
at the surface. We show that the width of the peak in $B_z$ strongly
depends on $z$ as in the inset of Fig.~\ref{jprho}, where the peak width
$W_B$ is defined by using the half-value radius; $B_z|_{\rho=W_B} =
(1/2)B_z|_{\rho=0}$. In the $s$-wave case,
the peak width does not strongly depends on $z$. On the contrary, it
depends significantly on $z$ in the $p_z$-wave case. At the surface,
$W_B$ is about four times larger than its value at $z=10 \xi$. 

\begin{figure}[tbp]
	\centering
	\includegraphics[width=0.98\columnwidth]{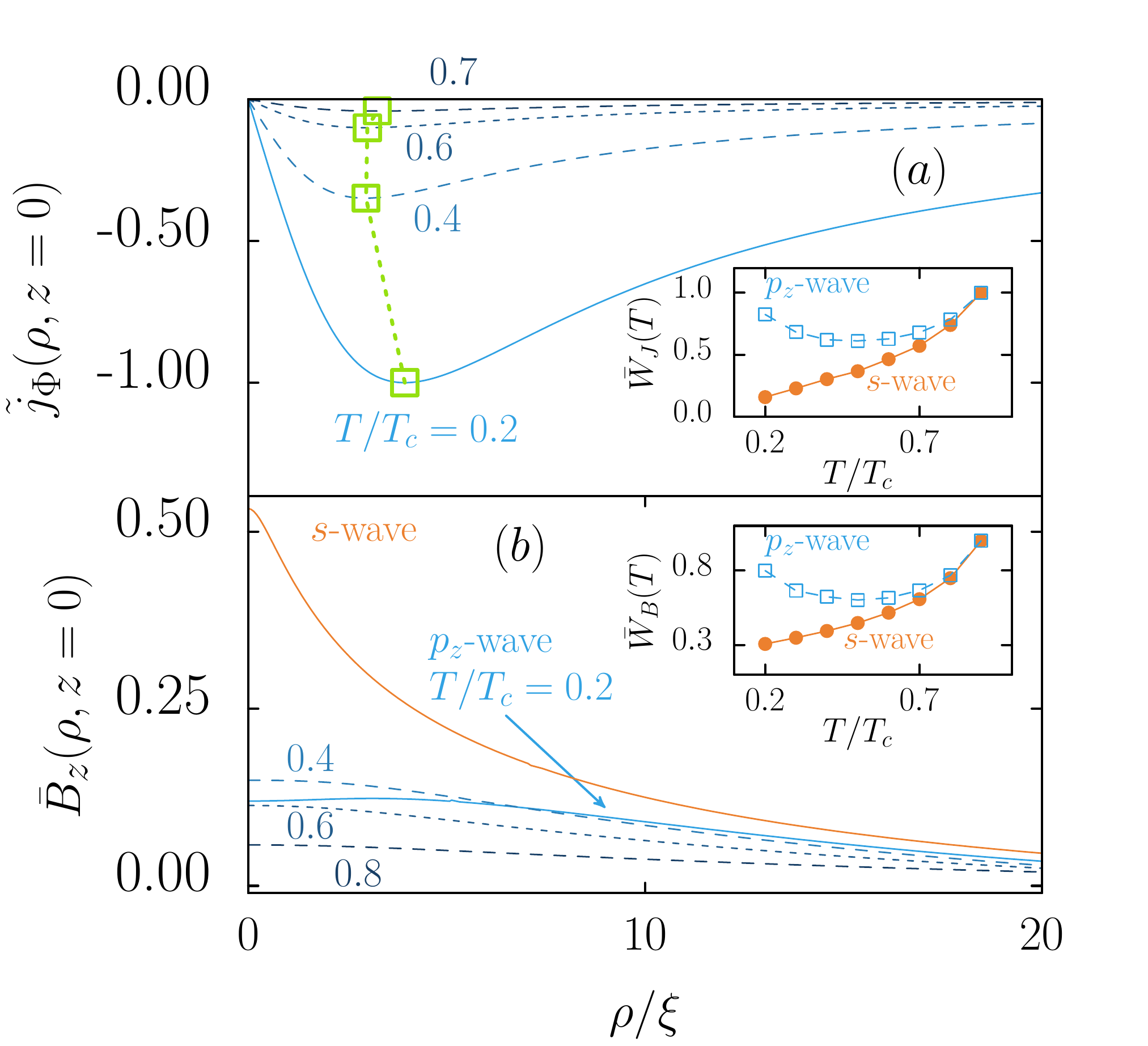}
	\caption{(a) Paramagnetic vortex-current at the surface of the $p_z$-wave SC. 
		The current density is normalized as $\tilde{j}=j_\Phi/|j_\Phi^{\mathrm{Min}}(T/T_c=0.2)|$
		with $j_\Phi^{\mathrm{Min}}(T/T_c=0.2)$ being the minimum value of $j_\Phi$ at $T/T_c=0.2$.
		The temperature is set to $T/T_c=0.2, 0.4, 0.6$ and $0.7$. 
		The open squares indicate the maximum value of 
		$|j_{\Phi}(\rho,z=0)|$ at each temperature. 
		(b) Magnetic field at the surface of the $p_z$- and $s$-wave SCs. 
		The temperature in the $p_z$-wave SC is set to $T/T_c=0.2, 0.4, 0.6$ and $0.8$, 
		and the temperature in the $s$-wave SC is $T/T_c=0.2$.
		Insets in (a) and (b) show the temperature dependences of $W_J$ and $W_B$, where $W_J$ is defined as $|j_\Phi(\rho=W_J,z=0,T)|=\max_{\rho}|j_\Phi(\rho,z=0,T)|$.
		We normalize $W_{J(B)}$ 
		as $\bar{W}_{J(B)}(T/T_c=0.8)=W_{J(B)}/W_{J(B)}(T/T_c=0.8)$ in each SC.
	}\label{fig:jpT}
\end{figure}

The temperature dependence of $j_\Phi$ is shown in Fig.~\ref{fig:jpT}(a),
where $j_\Phi$ is obtained at the surface of the $p_z$-wave SC. The
amplitude of the paramagnetic current increases with decreasing temperature. 
In particular, the paramagnetic current increases rapidly at low temperature. 
This low-temperature anomaly is characteristic to the 
paramagnetic response related to the odd-$\omega$ pairs induced by the
ABSs
\cite{Asano_PRL_2011,
Higashitani_PRL_2013, Higashitani_PRB_2014, Suzuki_PRB_2014, Suzuki_PRB_2015}.

\begin{figure*}[tbp]
	\centering
	\includegraphics[width=1.7\columnwidth]{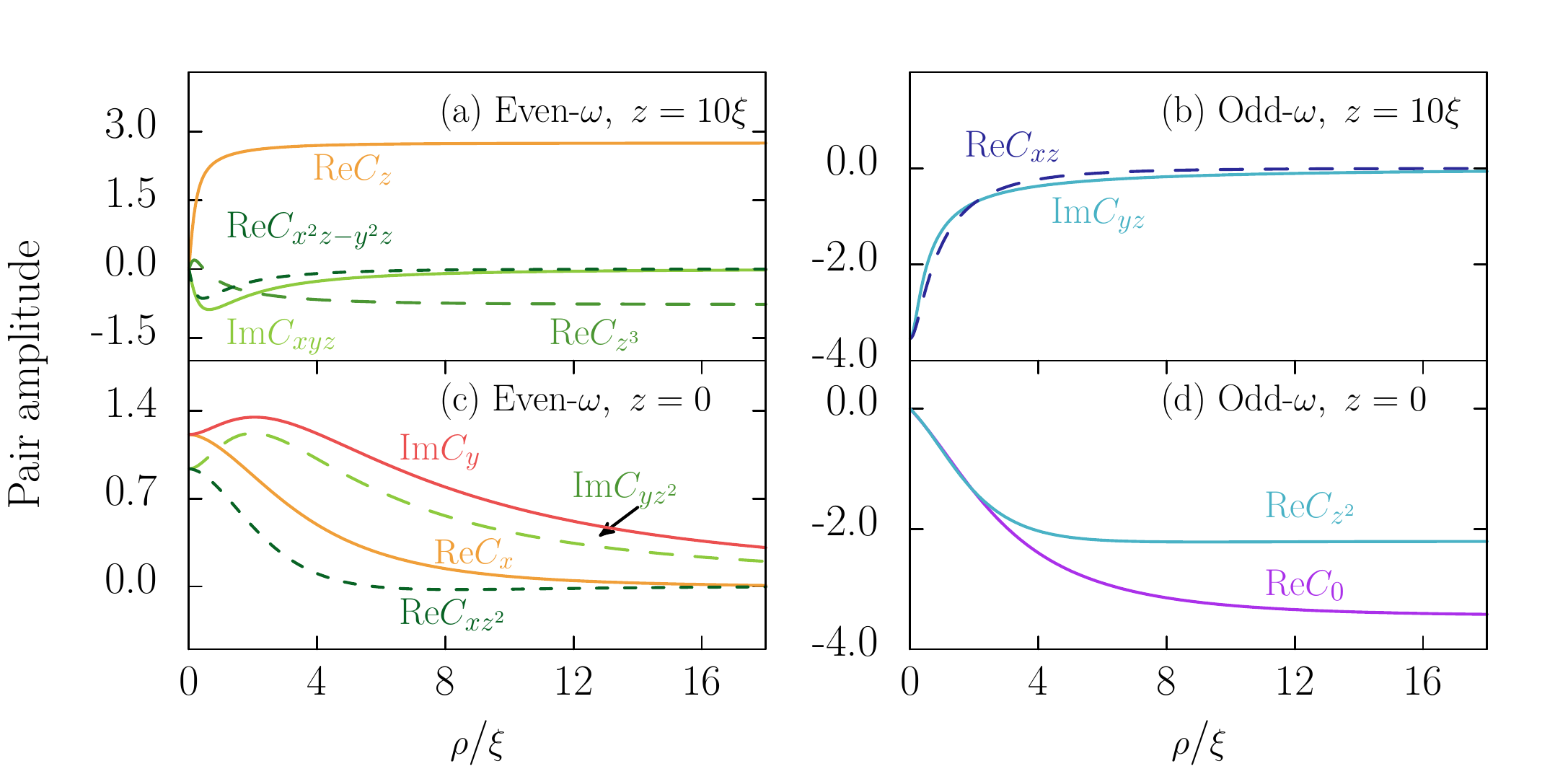}
	\caption{Pair amplitudes at (a,b) $z=10\xi$ and (c,d) $z=0$.
		The left (right) figures show the even frequency (odd
		parity) components.
		The parameters are set to $T=0.2T_c$, $\omega_n = \omega_0$, and $\Phi=0$.	
		The definition and notation of $C_{lm}(\bs{r})$ are shown 
		in Eq.~\eqref{ha_ex} and the Appendix \ref{sec:sphe_harm}.
		The even- and odd-frequency components are subdominant at $z=0$
		and $z=10\xi$ respectively. The other components are negligibly small. 
    }\label{fig:Fp}
\end{figure*}

\subsection{Symmetry conversion}	

We explore the origin of the paramagnetic circulating current in this subsection.
The magnetic response is related to the frequency symmetry of Cooper
pairs \cite{
Asano_PRL_2011, Yokoyama_PRL_2011, Higashitani_PRL_2013, 
Higashitani_PRB_2014, Suzuki_PRB_2014, Suzuki_PRB_2015, Asano_PRB_2015}. 
Namely, the direction of the vortex current is determined by
the spatial distribution of the even- and odd-$\omega$ Cooper pairs.
To analyze the symmetry of the Cooper pairs, we expand the pair
amplitudes at $\omega_0$ in terms of the real spherical harmonics (See
Appendix \ref{sec:sphe_harm} for the details): 
\begin{align}
	& f (\bs{r}, \bs{k}, \omega_0) = \sum_{lm} f_{lm}(\bs{r}, \bs{k}), 
	& f_{lm} = C_{lm}(\bs{r}) Y_{lm}(\bs{k}). 
	\label{ha_ex}
\end{align}
We refer to, for example, $f_{l=2,m=2}$ 
($d_{x^2-y^2}$-wave component) as $f_{x^2-y^2}$ for simplicity.
We hereafter fix $\Phi=0$ because of the rotational symmetry around
the vortex. 

Deep inside the $p_z$-wave SC, the symmetry conversion is caused only by 
the vortex singularity at $\rho=0$. In the homogeneous limit,
the pair amplitude has only the $p_z$-wave component: 
$ f = \Delta_\infty k_{z} / \sqrt{\omega_n^2+\Delta_\infty k_z^2 }
\sim f_z + f_{z^3}$ with $\Delta_\infty$ being the pair potential in
the homogeneous limit, where the denominator induces
non-$p_z$-wave component. At the vortex core, the pair amplitude
obtains an additional
phase from the phase winding of the vortex: $f \sim k_{z} (k_x + i
k_y) \sim f_{zx}+if_{yz}$ \cite{yokoyama_PRB_2008, tanuma_PRL_2009}.  In a spin-triplet SC, these
even-parity components should be an odd function of $\omega_n$ to
satisfy the Pauli exclusion principal (i.e., Berezinskii
rule\cite{Berezinskii}). Figure \ref{fig:Fp}(a) and
\ref{fig:Fp}(b) show the
symmetry conversion between the even- and odd-$\omega$ components at
$z=10\xi$. The even-$\omega$ $p_z$-wave component
is dominant at $\rho \gg \xi$, whereas only the odd-$\omega$
$d_{zx}+id_{yz}$-wave component has an amplitude at $\rho=0$. 

At the surface, the symmetry conversion is caused by the two factors: the
surface ABSs and vortex. The pairing symmetry is shown in
Figs.~\ref{fig:Fp}(c) and \ref{fig:Fp}(d). At $\rho \gg \xi$, the pair amplitude can be
written as a superposition of the $s$-wave and $d$-wave pairings
\cite{tanaka_PRB_2007, Suzuki_PRB_2014, Suzuki_PRB_2015}: 
$	f \sim f_0 + f_{z^2}$. 
Namely, only the odd-$\omega$ pairs are present at the surface far
from a vortex. At the intersection of the vortex core and surface
(i.e., $\rho=z=0$), 
these odd-$\omega$ pairs
 have an additional phase from the vortex. 
As a result, the $p_x + i p_y$-wave and $f_{xz^2} + i f_{yz^2}$
 are dominant as shown in Fig.~\ref{fig:Fp}(c). Note that
these even-$\omega$ pairs are generated from the odd-$\omega$ pairs, 
and have different pairing symmetry from the pair potential. 

The relation between 
the frequency-symmetry conversion and supercurrent inversion can be
confirmed by analyzing the current density $\bs{j}$ in Eq.~\eqref{jeq}. 
Close to the vortex core (i.e., $\rho \ll \xi$), $\bs{j}$
 can be expressed in terms of the pair amplitudes
\cite{Higashitani_JPSJ_2014, Suzuki_PRB_2014, Suzuki_PRB_2016}. 
Specifically, it
can be expressed by the even-$\omega$ (odd-$\omega$) component 
$f_{e(o)}$: 
\begin{align}
	& j_\Phi(\bs{r})
	= -j_0
	\frac{T}{T_c}
	\sum_{n=0}^{N_{\mathrm{c}}}
	\int\frac{d\Omega}{4\pi}
	\sin\theta\sin\phi\mathrm{Im}g(\hat{\bs{k}},\bs{r},\omega_n), 
  \label{eq:cur_freq1}
	\\
	& g=\sqrt{1-|f_e|^2+|f_o|^2-2i\mathrm{Im}f_of_e^*}, 
  \label{eq:cur_freq2}
	\\
  & f_{e(o)} = \frac{1}{2}[f(\bs{r}, \bs{k},  \omega_n) 
                    	+(-) f(\bs{r}, \bs{k}, -\omega_n)]
\end{align}
where we have used the normalization condition (i.e., $g^2=1-s_\mu f\ut{f}$)
and $j_0={4\pi N_0|e|v_F T_c}$. 
In the Appendix~\ref{sec:ConvFreq}, applying the conversion $f_e
\leftrightarrow f_o$ to Eqs.~\eqref{eq:cur_freq1} and
\eqref{eq:cur_freq2}, we have shown that the frequency-symmetry
conversion by the surface ABSs results in the vortex-current
inversion.

\begin{figure*}[tb]
	\centering
	\includegraphics[width=1.8\columnwidth]{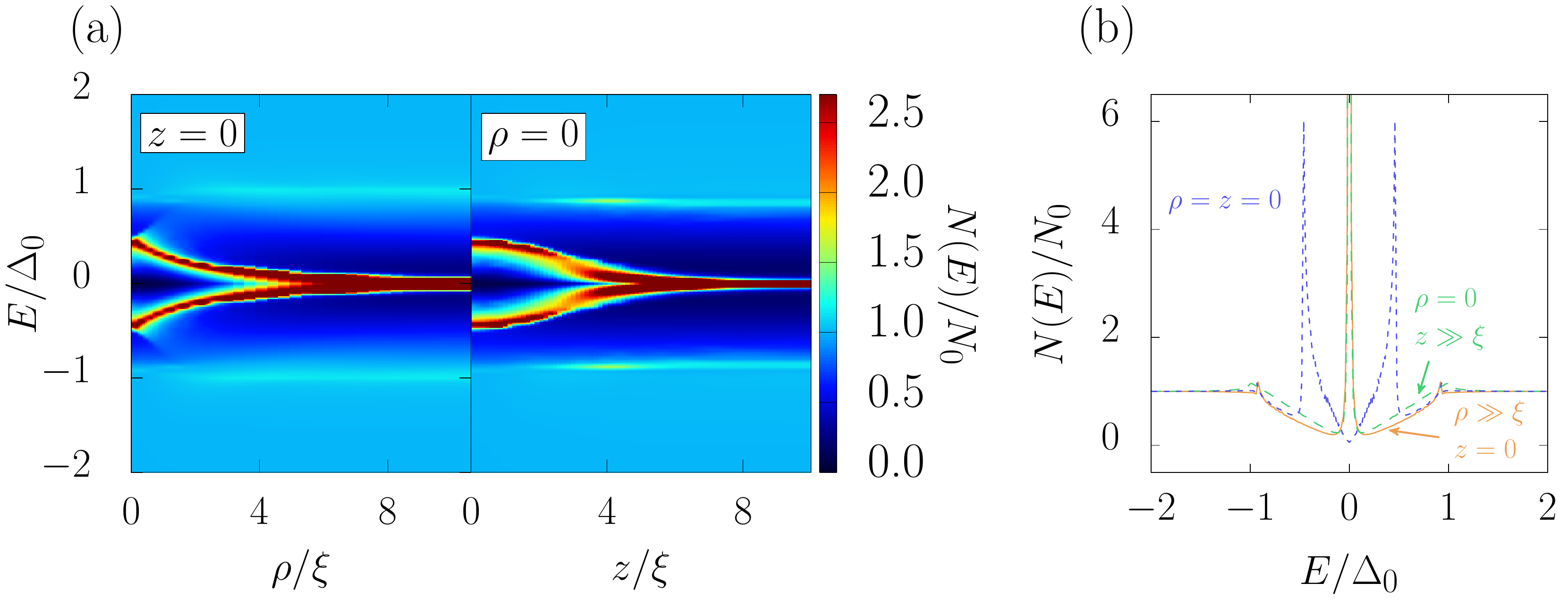}
	\caption{(a) Contour plots of
	the LDOS at the surface and the vortex core. (b) Comparison of LDOS. The
	parameters are set to the same values used in Fig.~\ref{fig:cur}. The
	smearing factor is $\delta = 0.005$. }
	\label{fig:ldosp}
\end{figure*}

\subsection{Local density of states}

Before discussing the obtained results, we quickly overview the in-gap
states related to this system: surface ABS and CdGM mode. The
$p_z$-wave SC hosts the surface ABSs\cite{hara_PTP_1986} at its surface perpendicular to
the $z$-direction because of the quasiparticle
interference
\footnote{At a surface of an SC, the incoming and outgoing
quasiparticles interfere.  These two quasiparticles feel, in general,
different pair potentials depending on the momentum:
$\Delta(k_z,\bs{k}_\parallel)$ and $\Delta(-k_z,\bs{k}_\parallel)$
where the surface is perpendicular to the $z$ axis. In the $p_z$-wave
SC, the phase difference between these two pair
potentials is $\pi$:
$\Delta(k_z,\bs{k}_\parallel) = -\Delta(-k_z,\bs{k}_\parallel)$. The
$\pi$-phase shift results in forming the surface bound state at the zero
energy.}. In the $p_z$-wave case, the ABSs form the flat-band
zero-energy surface states\cite{hu_PRL_1994, kashiwaya_RPP_2000, sato_PRB_2011}. 
The CdGM modes are
quasiparticle states bounded at a vortex core\cite{caroli_PL_1964}. Because of the
confinement by the local suppression in the pair potential, the in-gap
quasiparticle states appears at the vortex core \footnote{The energy
levels for the CdGM modes are given by $E=\pm (n-1/2) \Delta_\infty^2/E_F$.
In the quasiclassical regime, the lowest CdGM mode seem to sit at
$E=0$ because $\Delta/E_F \ll 1$ is assumed in 
the quasiclassical theory. }. 
In the Appendix \ref{sec:num_res}, we show the quasiparticle spectra of the
$s$-wave SC with a vortex as a reference, where we explain the spatial
distribution of the CdGM modes. 

The LDOS at the surface of the $p_z$-wave SC are shown in
Fig.~\ref{fig:ldosp}, which can be measured by, for example,
a scanning tunnel spectroscopy measurement. 
The flat-band zero-energy ABSs appear at far from the vortex core
(i.e., $\rho \gg \xi$). Approaching to the vortex, the bound-states
energy is lifted from $E=0$ and the zero-energy peak is split into two
peaks with $E \approx 0.5 \Delta_\infty$ at $\rho=0$.
The evolution of the CdGM modes at the lowest energy are shown in 
Fig.~\ref{fig:ldosp}, where we fix $\rho = 0$. Deep inside the SC
($z \gg \xi$),
the CdGM mode stay at $E \approx 0$. However, with approaching to the
surface, the zero-energy peak splits into two peaks. 

We conclude the 
zero-energy peak splitting is a result of the interference between the
surface ABSs from the $p_z$-wave nature and the CdGM modes from the
quantum vortex: the quantum states staying at $E=0$
repel each other in energy space due to the superposition in
real space. 

The splitting of the LDOS at $E=0$ can also be interpreted in another way. 
The ABS appears at the zero-energy when 
$\Delta(\bs{k}_{\mathrm{in}}) = 
-\Delta(\bs{k}_{\mathrm{out}})$ during the reflection at a surface,
where $\bs{k}_{\mathrm{in}}$ ($\bs{k}_{\mathrm{out}}$) is the incoming
(outgoing) momentum. A similar sign change occurs when the
quasiparticle passes through the vortex core. 
However, 
on the quasiclassical paths passing through the origin (intersection
of the surface and the vortex core), this sign change never
occurs. The sign change from the reflection and that from the vortex
core compensate each other. 
Therefore, the zero-energy state does not
appear at the origin.

\section{Kramer-Pesch theory}\label{resultanasec}

We apply the Kramer-Pesch
approximation\cite{kramer_ZP_1974, nagai_JPSJ_2006,
nagai_JPSJ_2019} to more general superconducting states and
generalize our results obtained
numerically;  
the vortex-current reversal and disappearance
of the zero-energy peak at $z=\rho=0$ in the LDOS. 
The Kramer-Pesch approximation,
which is valid when $k_B T \ll |\Delta_\infty|$ and $\rho \ll \xi$, 
allow us to obtain the analytic solution of the quasiclassical Green's
function.
We assume that the gap function is written as
\begin{align}
	\Delta(\bs{r},\hat{\bs{k}})
	=h(\rho,z)e^{-i\Phi}\Psi(\hat{\bs{k}}),\label{KP_delta}
\end{align}
where $h(\rho,z)$ is a real function describing the spatial
profile of $\Delta$ and 
 $\Psi(\hat{\bs{k}})$ is the momentum dependence of $\Delta$ which
 satisfies
 $\Psi(\bs{k}_\parallel,k_z)=s_\theta\Psi(\bs{k}_\parallel,-k_z)$ with
 $\bs{k}_{\parallel}=(k_x,k_y,0)$ and $s_\theta=\pm1$.
Surface ABSs are formed when $s_\theta=-1$.
In the Kramer-Pesch approximation, the azimuthal components of current density 
$j_\Phi$ for each $s_\theta$ are written as 
\begin{widetext}
\begin{align}
	j_\Phi(\bs{r})|_{s_\theta=1}&\simeq 4\pi N_0|e|v_F\int\frac{d\Omega}{4\pi}\sin^2(\phi-\Phi)\frac{C^{-1}}{4k_B T}\rho\tilde{D}\label{ksj},\\
	 j_\Phi(\bs{r})|_{s_\theta=-1}
	&\le4\pi N_0|e|v_F
	\int\frac{d\Omega}{4\pi}\sin^2(\phi-\Phi)
	\left[-\frac{C^{-1}}{8k_B T}+\frac{2}{C^{-1}}\right]
	\rho\tilde{D}
	\label{kpj},\\
	\tilde{D}(\rho,\Phi,\bs{k})&=\frac{2}{v_FC(\rho,\Phi,\bs{k})}\int^{\infty}_{-\infty}d\sigma^\prime\frac{1}{\sigma^\prime} 
		   h\left(|\sigma|\sin\theta,\left|\sigma\cos\theta-\rho\frac{\cos(\phi-\Phi)}{\tan\theta}\right|\right)
		\exp\left[-2|\Psi(\bs{k})|\bar{F}(\rho,\Phi,\sigma^\prime,\bs{k})\right],\\
  C(\rho,\Phi,\bs{k})&=\frac{2}{v_F}\int^{\infty}_{-\infty}d\sigma^\prime\exp\left[-2|\Psi(\bs{k})|\bar{F}(\rho,\Phi,\sigma^\prime,\bs{k})\right],\\
  \bar{F}(\rho,\Phi,\sigma,\hat{\bs{k}})&=\int_{\rho\cos(\phi-\Phi)(\sin\theta)^{-1}}^{\sigma}\mathrm{sign}(\sigma^{\prime})h(|\sigma^{\prime}|\sin\theta,|\sigma^{\prime}\cos\theta+z_0|)d\sigma^{\prime}\label{barFeq},
\end{align}
\end{widetext}
where $\tilde{D} \ge 0$ and $C \ge 0$. 
The detailed derivation of Eq.~\eqref{kpj} is
explained in the Appendix \ref{sec:kp_app}.
The conventional vortex current has a positive value as
in Eq.~\eqref{ksj}.
On the other hand, Eq.~\eqref{kpj} shows that the current density can
be negative at low temperature because of the first term. 
Thus, the vortex current on the surface with the ABS is reversed. 

In the Kramer-Pesch approximation, the surface DOS for $s_\theta=-1$ is written as
\begin{align}
	\frac{N(E,\rho,z=0)}{N_0}&=\int\frac{d\Omega}{4\pi}\frac{(\rho D+E)^2}{(\rho D+E)^2+C^{-2}}\le1,\label{kpldos}\\
	D&=\frac{\sin(\phi-\Phi)}{\sin\theta}\tilde{D}.
\end{align}
Equation \eqref{kpldos} shows that the surface DOS is less than
$N_0$ and becomes zero at $\rho=E=0$ (see the Appendix
\ref{sec:kp_app} for the details.).
Therefore, the zero-energy peak in the LDOS
at $\rho=z=0$ disappears when $s_\theta=-1$.

Applying the Kramer-Pesch approximation, we
demonstrate the current
inversion and the disappearance of the zero-energy peak in the LDOS under
the condition $\Delta(k_x, k_y, k_z) = -\Delta(k_x, k_y, -k_z)$. 
This indicates that our numerical results are not
unique only to the $p_z$-wave SC. The vortex
supercurrent inversion can occur in superconductors with flat-band surface ABSs.

\section{Conclusion}\label{conclusionsec}

We have demonstrated that the local frequency-symmetry conversion of
Cooper pairs inverts
the direction of the vortex supercurrent. Using the quasiclassical
Eilenberger theory, a quantum vortex penetrating a surface hosting the
ABSs both numerically and analytically has been investigated. In the
self-consistent simulations, comparing the vortex supercurrents in the
$p_z$-wave and $s$-wave SCs, we have found that the supercurrent flows
in the opposite direction near the surface of the $p_z$-wave SC.  By
analyzing the anomalous Green's function, the frequency-symmetry
conversion among Cooper pairs by ABSs has been shown to trigger the
supercurrent inversion.  In the analytic calculation with the
Kramer-Pesch approximation, the condition of the vortex current
inversion has been generalized. 

From the quasiparticle spectra obtained from the Green's function, we
have found that the zero-energy states from the ABSs and CdGM mode
are coupled with each other and shift to non-zero energies when the
current inversion occurred. Within the Kramer-Pesch approximation, 
it has been analytically shown that the LDOS is zero at the
intersection of the vortex core and surface because of the
interference between the ABSs and CdGM mode.

\begin{acknowledgments}
The authors are thank A.~A.~Golubov for the fruitful discussions. 
S.~Y. would like to take this opportunity to thank the ``Nagoya University
Interdisciplinary Frontier Fellowship'' supported by JST and Nagoya University.
S.-I.~S. is supported by JSPS Postdoctoral Fellowship for Overseas
Researchers and a Grant-in-Aid for JSPS Fellows (JSPS KAKENHI Grant
No. JP19J02005). 
This work was supported by Scientific Research (A)
(KAKENHI Grant No. JP20H00131), and Scientific Research
(B) (KAKENHI Grant No. JP20H01857). 
\end{acknowledgments}

\appendix

\section{Maxwell equation, Riccati parametrization, and boundary conditions}\label{sec:form}

In this section, we introduce the formulations of the vector
potential, Riccati parametrization, and boundary conditions.
Under the rotational symmetry around the $z$ axis, the magnetic field $\bs{B}$ and vector
potential $\bs{A}$ are obtained from $\bs{j}$ as 
\begin{align}
	\bs{B}(\bs{r})&=\frac{1}{c}\int
  \bs{j}(\bs{r}^\prime)\times
  \frac{\bs{r}-\bs{r}^\prime}{\left|\bs{r}-\bs{r}^\prime\right|^3}
	dV^\prime,\label{B_form}\\
	\bs{A}(\bs{r})&=
	\bs{e}_\Theta\int_z^{\infty}B_\rho(\rho,z^\prime)dz^\prime
	\notag\\
	& \hspace{6mm}+\bs{e}_\Theta\frac{1}{\rho}\int_{0}^\rho\rho^\prime B_z(\rho^\prime,z=\infty)d\rho^\prime.\label{A_form}
\end{align}
We introduce a dimensionless Maxwell equation \eqref{maxeq} for numerical
calculations:
\begin{align}
	\bs{\nabla}^\prime \times 
	\left(\bs{\nabla}^\prime\times\bs{A}^\prime\right)=\frac{6}{\kappa^2}\bs{j}^\prime,
	\label{maxeq_kappa}
\end{align}
where $\bs{\nabla}'= \xi\bs{\nabla}$, $\bs{A}$ and $\bs{j}$ are normalized as 
$\bs{A}^\prime=(2\pi\xi/\Phi_0)\bs{A}$ 
and $\bs{j}^\prime=\bs{j}/j_0$ with $\Phi_0=\pi \hbar c/|e|$ and $j_0=4\pi N_0|e|v_FT_c$.

In order to solve the Eilenberger equation~\eqref{eilenberger}
numerically, we
introduce the coherence function\cite{schopohl_PRB_1995,
schopohl1998transformation} $\gamma$ as
\begin{align}
	g&=\frac{1-s_\mu\gamma\ut{\gamma}}{1+s_\mu\gamma\ut{\gamma}},\quad
	f=\frac{2\gamma}{1+s_\mu\gamma\ut{\gamma}}.\label{coh_fun}
\end{align}
Substituting Eq.~\eqref{coh_fun} into Eq.~\eqref{eilenberger}, we obtain the 
Riccati-type differential equations:
\begin{align}
	\hbar\bs{v}_F\cdot \bs{\nabla} \gamma-
	2\left(\omega_n-\frac{ie}{c}\bs{v}_F\cdot\bs{A}\right)
	\gamma-\Delta+\Delta^*\gamma^2=0,\label{riccati}\\
	\hbar\bs{v}_F\cdot \bs{\nabla} \ut{\gamma}+
	2\left(\omega_n-\frac{ie}{c}\bs{v}_F\cdot\bs{A}\right)
	\ut{\gamma}+s_\mu\Delta^*-s_\mu\Delta\ut{\gamma}^2=0.\label{ut_riccati}
\end{align}
The solutions of Eq.~\eqref{eilenberger} are
given by solving Eq.~\eqref{riccati} and \eqref{ut_riccati} along the quasiclassical paths and substituting the obtained $\gamma$ and $\ut{\gamma}$ into Eq.~\eqref{coh_fun}.
We solve Eqs.~\eqref{gapeq}, \eqref{A_form}, \eqref{riccati} and
\eqref{ut_riccati} numerically in a self-consistent manner. 
The coherence functions $\gamma$ and $\ut{\gamma}$ in the
homogeneous limit
(i.e., $\nabla\cdot\gamma=0$) are given by
\begin{align}
	\gamma&=\frac{\Delta_\infty}{\omega_n+\mathrm{sign}(\omega_n)\sqrt{\omega_n^2+|\Delta_\infty|^2}},\label{coh_fun_bulk1}\\
	\ut{\gamma}&=\frac{s_\mu\Delta_\infty^*}{\omega_n+\mathrm{sign}(\omega_n)\sqrt{\omega_n^2+|\Delta_\infty|^2}},\label{coh_fun_bulk2}
\end{align}
where $\Delta_\infty$ is the bulk value of $\Delta$.

We impose the boundary conditions on the coherence
functions. 
(I) The coherence functions are equal to the values in
Eqs.~\eqref{coh_fun_bulk1} and \eqref{coh_fun_bulk2} 
deep inside the SC (i.e., $\rho\gg\xi,~z\gg\xi$).
(II) At the surface (i.e., $z=0$), we impose
\begin{align}
	\gamma(\hat{\bs{k}}_{\parallel},\hat{\bs{k}}_{z},\bs{r}_{\parallel},z=0,\omega_n)=\gamma(\hat{\bs{k}}_{\parallel},-\hat{\bs{k}}_{z},\bs{r}_{\parallel},z=0,\omega_n),\label{bc}
\end{align}
which means the specular reflection
where $\bs{r}_{\parallel}=(x,y)$.
As a result of Eq.~\eqref{bc}, 
the pair amplitude $f$ is forced to be an even
function with respects to the $k_z$ inversion at the surface.

\begin{figure}[tb]
	\includegraphics[width=\linewidth]{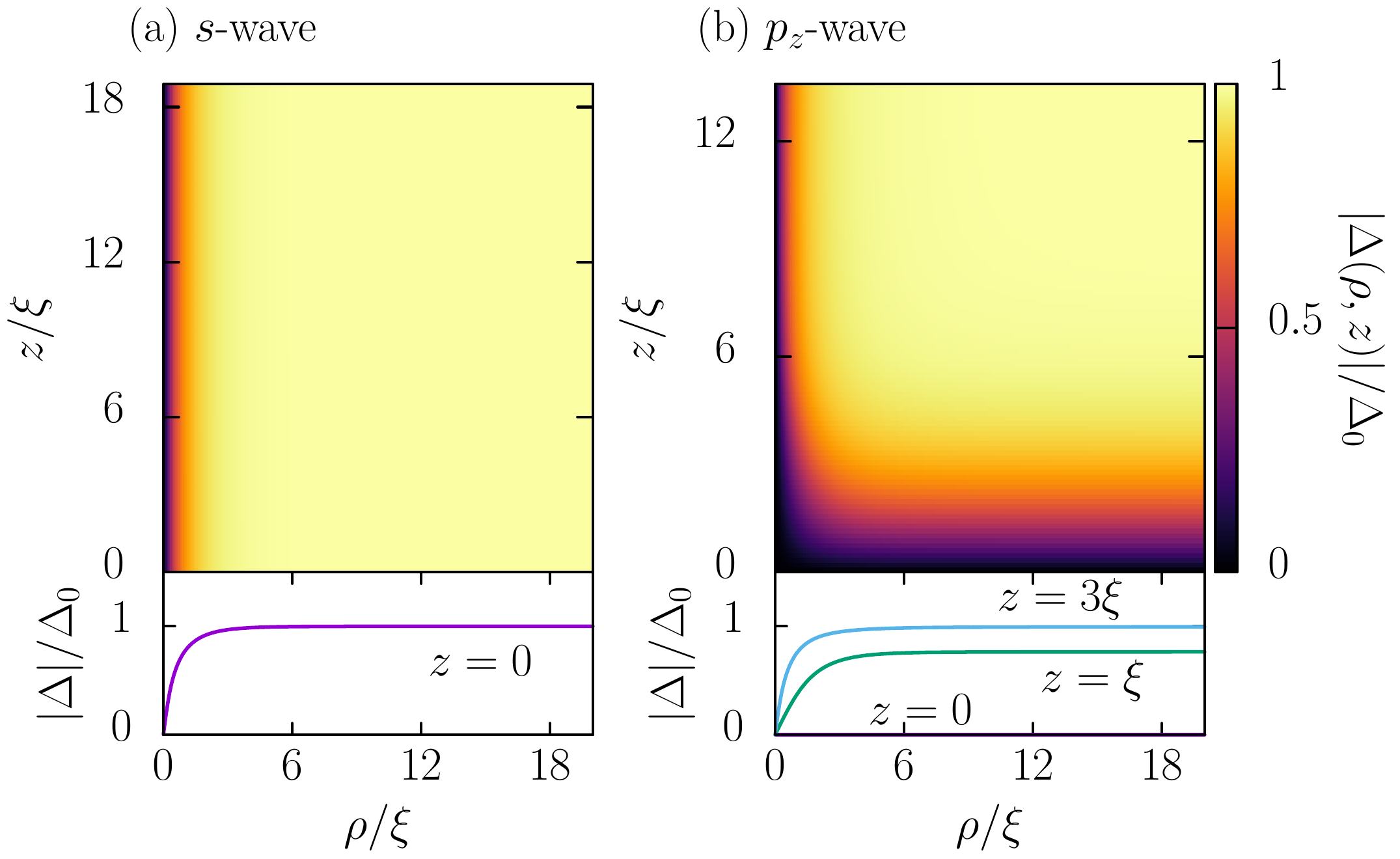}
	\caption{Spatial dependences of 
		$|\Delta|$ in the (a) $s$-wave and (b) $p_z$-wave SCs.
	The pair potential is normalized by
	its bulk value.
	The parameters are set to $T/T_c=0.2$, $\Omega_{c}=40k_BT_c$ and $\kappa_{GL}=6\sqrt{6}$.  }
	\label{gap}
\end{figure}
\begin{figure}[tb]
	\centering
	\includegraphics[width=0.48\textwidth]{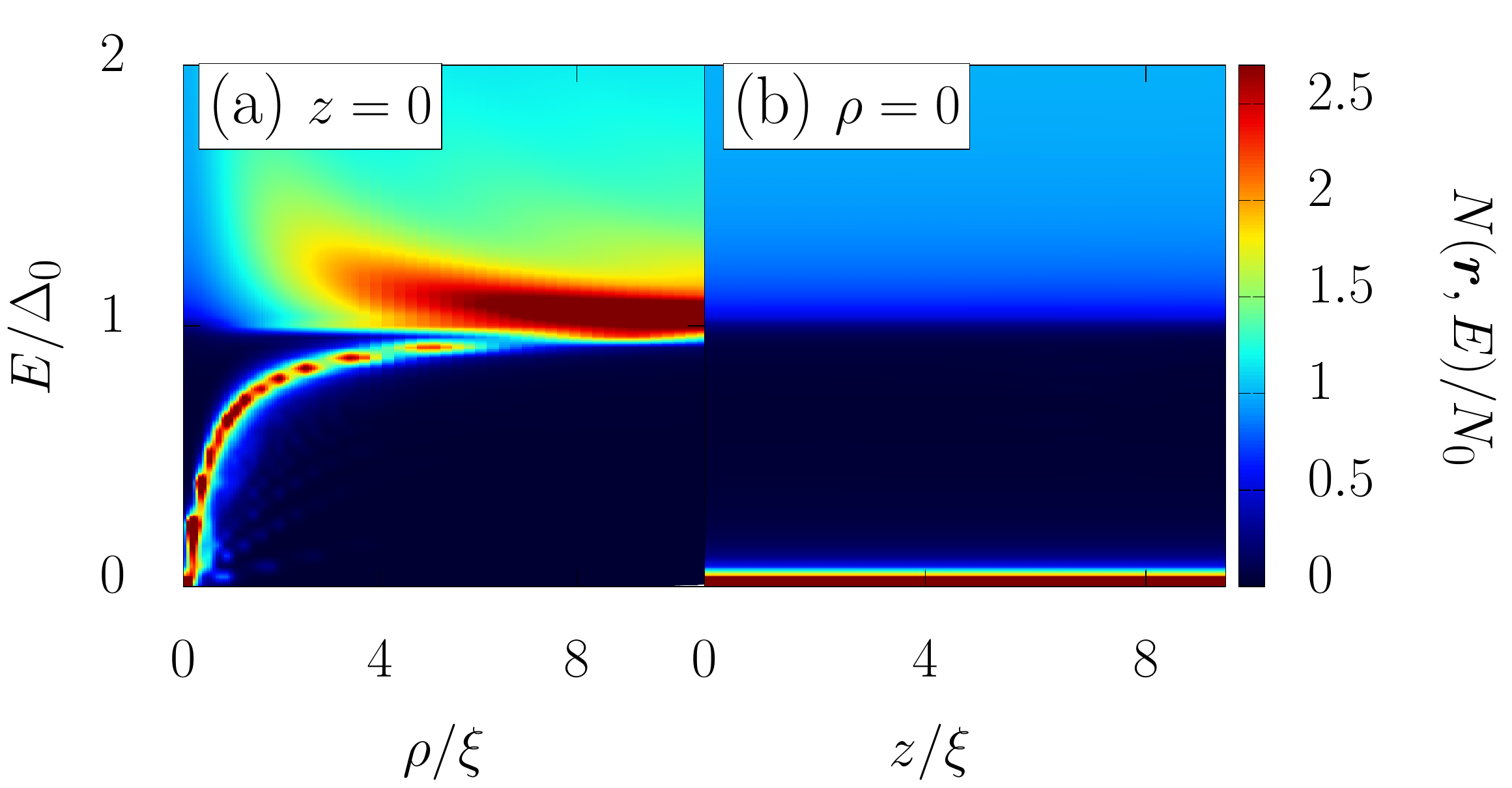}
	\caption{Contour plots of the LDOS
	at (a) surface and (b) vortex core in the $s$-wave SC.
	The parameters are the same as in Fig.~\ref{gap}.
	The smearing factor is $\delta=0.005$.}
	\label{ldos_s}
\end{figure}

\section{Additional numerical results } \label{sec:num_res}

In this section, we show the spatial dependence of the pair potential 
and local density of state (LDOS) for the $s$-wave SC.
We show the spatial dependences of the pair
potential in Fig.~\ref{gap}. 
As show in Fig.~\ref{gap}(a), the pair potential in
the $s$-wave SC is
suppressed near the vortex core but not at the surface. 
On the other hand, the pair function in the $p_z$-wave SC
disappears at the
vortex core and the surface as shown in Fig.~\ref{gap}(b).

In Fig.~\ref{ldos_s}, we show the LDOS in the $s$-wave SC. 
Figures~\ref{ldos_s}(a) and (b) show the LDOS at the
surface and vortex core.
In Fig.~\ref{ldos_s}, there exist sub-gap bound states near the core
known as the Caroli-de-Gennes-Matricon (CdGM) modes. 
The approximate energy dispersion of the CdGM modes have been derived 
as $E_n\simeq (n-1/2)\Delta_\infty^2/E_F$ in Ref.~\onlinecite{caroli_PL_1964}, where the half integer $(n-1/2)$ is the quantum number of the angular momentum.
The eigenfunction of the CdGM mode labeled by $n$ has 
a maximum amplitude at a radius distance of the order $\rho\simeq (n-1/2)/k_F$\cite{gygi_PRB_1991}.
Thus, the evolution of sub-gap states in Fig.~\ref{ldos_s}(a)
corresponds to the dispersion relation 
of the CdGM modes.

\section{Renormalized spherical harmonics}\label{sec:sphe_harm}
We use the real spherical harmonics $Y_{lm}$ to analyze
the pair amplitudes. 
We abbreviate the expansion coefficients $f_{lm}$ in the main texts for simplicity.
We summarize $Y_{lm}$ and the notation of $f_{lm}$ in Table \ref{table:fnotation}.
The spherical harmonics
$Y_{lm}$ is normalized to satisfy the orthonormal relation
\begin{align}
	\int Y_{lm}Y_{l^\prime m^\prime}d\Omega=\delta_{ll^\prime}\delta_{mm^\prime},
\end{align}
where $d\Omega=\sin\theta d\phi d\theta$.

\begin{table*}[tb]
	\caption{Notation of the expansion coefficients $f_{lm}$}
  \label{table:fnotation}
  \centering
	\includegraphics[width=0.7\textwidth]{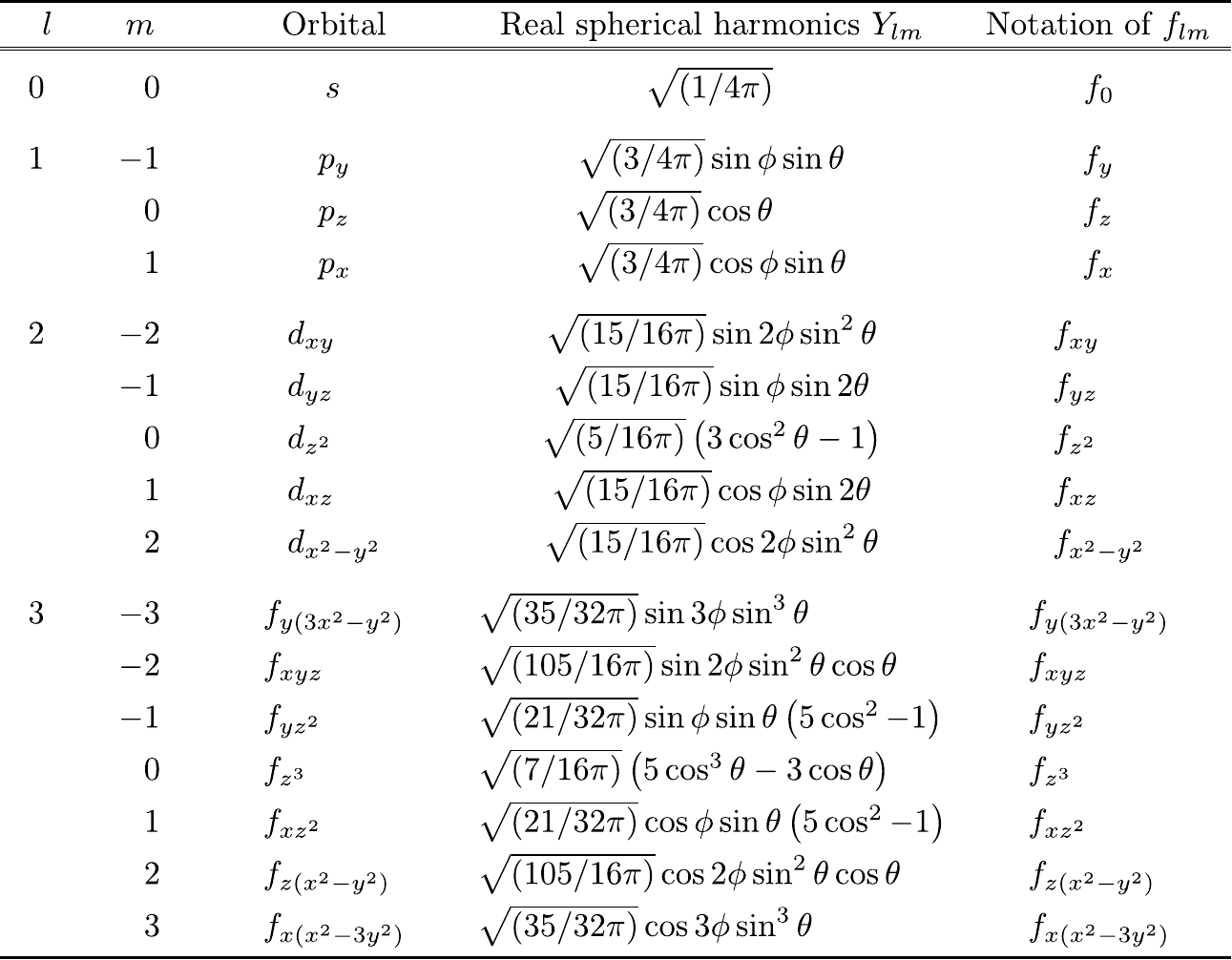}
\end{table*}

\section{Vortex supercurrent and frequency symmetry of Cooper pairs} \label{sec:ConvFreq}

The direction of the vortex supercurrent is related to the frequency
symmetry of Cooper pairs. 
The azimuthal component of the current density \eqref{jeq} is
\begin{align}
	& j_\Phi(\bs{r})
	= -j_0
	\frac{T}{T_c}
	\sum_{n=0}^{N_{\mathrm{c}}}
	\int\frac{d\Omega}{4\pi}
	\sin\theta\sin\phi\mathrm{Im}g(\hat{\bs{k}},\bs{r},\omega_n), 
\end{align}
where $j_0={4\pi N_0|e|v_F T_c} $.
Namely, the direction of the current is determined by the sign of 
$\mathrm{Im}[g]$. 
Using the normalization condition for the quasiclassical Green's
function (i.e., $g^2=1-s_\mu f\ut{f}$), we have 
\begin{align}
	& g=\sqrt{1-|f_e|^2+|f_o|^2-2i\mathrm{Im}f_of_e^*}, 
	\label{eq:g}
	\\
	& f_{e(o)} = \frac{1}{2}[f(\bs{r}, \bs{k}, \omega) +(-) f(\bs{r}, \bs{k}, -\omega)], 
\end{align}
where $f_e$ and $f_o$ represent respectively the even- and odd-frequency
component, and we have used the symmetry relation between $f$ and $\ut{f}$
[i.e., $\ut{f}(\bs{r}, \bs{k}, \omega)=f^*(\bs{r},-\bs{k}, \omega)$].

\begin{figure}[tb]
	\centering
	\includegraphics[width=0.46\textwidth]{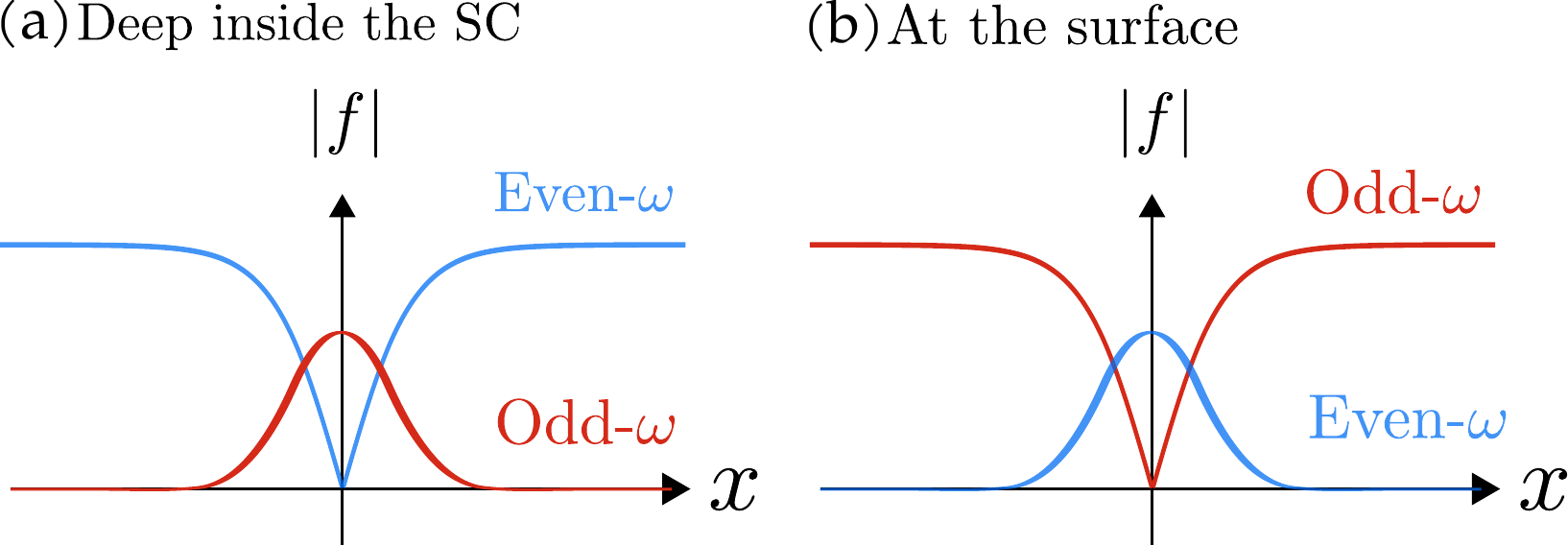}
	\caption{Schematic picture of the spatial distribution of the pair
	amplitudes around the vortex in the $p_z$-wave SC.  The even and
	odd-$\omega$ pairs are dominant near the vortex core at the surface
  and deep inside the SC.}
	\label{pair_amp}
\end{figure}

We analyze the direction of the vortex supercurrent in the vicinity of
the vortex core. From Fig.~\ref{pair_amp}, deep inside
the SC, the pair amplitudes around the vortex core can be approximated
as 
\begin{align}
	f_e \approx \rho F, \quad 
	f_o \approx -RF e^{-i\phi},
	\label{eq:even-odd}
\end{align}
where $F=F(\hat{\bs{k}})$ and we have introduced the constant $R$
($0<R<1$) that expresses the relative amplitude of $f_o$ to $f_e$. 
The subdominant component has an additional phase factor ($-e^{-i\phi}$)
from the phase winding of the vortex. 
Substituting Eq.~\eqref{eq:even-odd} into Eq.~\eqref{eq:g}, we have
\begin{align}
	g&=\sqrt{1-\rho^2|F|^2+R^2|F|^2+2i\rho R|F|^2\mathrm{Im}e^{-i\phi}}\\
   &=\sqrt{1+R^2|F|^2}
	 -i\rho\frac{R|F|^2\sin\phi}{\sqrt{1+R^2|F|^2}}
	 +O(\rho^2)
\end{align}
with $\rho \ll \xi$. 
The equation above results in 
\begin{align}
  -j_0\sin\phi\mathrm{Im}g
	=
	\rho\frac{j_0R|F|^2\sin^2\phi}{\sqrt{1+R^2|F|^2}}+O(\rho^2), 
	\label{eq:cur_dir1}
\end{align}
meaning that the electric current flows in the opposite direction of
the phase winding $e^{-i\Phi}$ around the vortex core (i.e., $j_\Phi >0$). 

At the surface of a $p_z$-wave SC, the symmetry conversion occurs
because of the ABSs. The pair amplitudes can be expressed as 
\begin{align}
	f_e=-RF e^{-i\phi},\quad f_o=\rho F, 
	\label{eq:even-odd2}
\end{align}
where we can show $R^2|F|^2=1$ by solving the Riccati-type differential equations \eqref{riccatieq} at $T=0$ and $\rho=0$. 
	We assume $f_e$ near the core has the maximum value at $\omega_n=0$.
	From this assumption, we obtain $R^2|F|^2<1$ at finite matsubara frequencies.
Substituting Eq.~\eqref{eq:even-odd2} into Eq.~\eqref{eq:g}, we have
\begin{align}
	g&=\sqrt{1-R^2|F|^2+\rho^2|F|^2+2i\rho R|F|^2\mathrm{Im}e^{i\phi}}
       \\&=\sqrt{1-R^2|F|^2}
			 +i\rho\frac{R|F|^2\sin\phi}{\sqrt{1-R^2|F|^2}}+O(\rho^2). 
\end{align}
Therefore, the direction of the current density is given by 
\begin{align}
  -j_0\sin\phi\mathrm{Im}[g]
	=-\rho\frac{j_0R|F|^2\sin^2\phi}{\sqrt{1-R^2|F|^2}}+O(\rho^2)
	\label{eq:cur_dir2}
\end{align}
Comparing Eqs.~\eqref{eq:cur_dir1} and \eqref{eq:cur_dir2}, we see
that the frequency-symmetry conversion results in the vortex-current inversion.

\section{Analytical Calculations by Kramer-Pesch approximation} \label{sec:kp_app}
\begin{figure}[tb]
	\centering
	\includegraphics[width=0.2\textwidth]{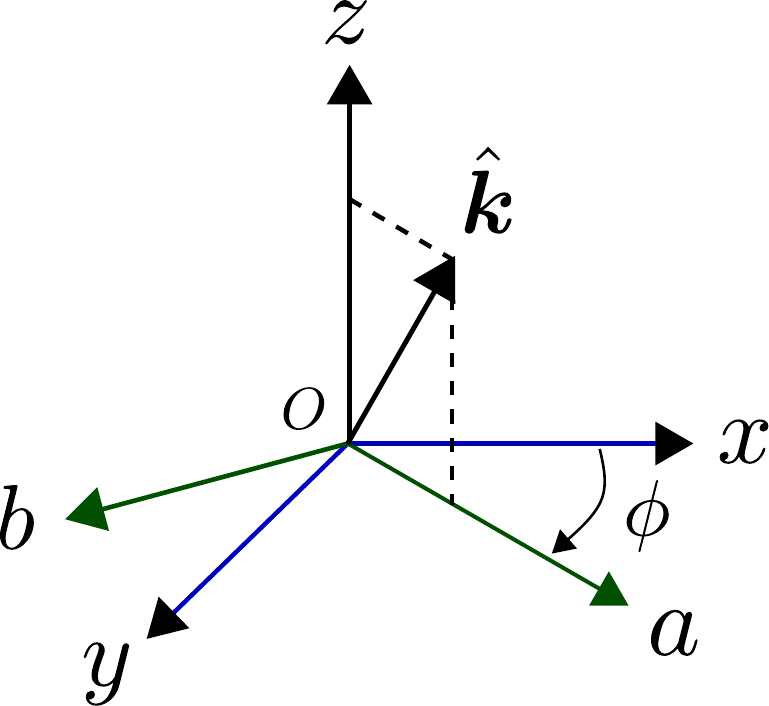}
\caption{Coordinate transformation from $(x,y,z)$ to $(a,b,z)$.}
	\label{fig_cotr}
\end{figure}

In this section, we analytically calculate the quasi-classical Green's
functions by the Kramer-Pesch approximation
\cite{kramer_ZP_1974,nagai_JPSJ_2006, nagai_JPSJ_2019} and
reproduce two
numerical results, the reversal of the vortex current 
and suppression of the LDOS at the intersection point 
of the surface and vortex core, for more general SCs.  
In the Kramer-Pesch approximation, we obtain the solutions of the Riccati-type differential equations up to
the first order of the displacement $b$ from the vortex core and Matsubara frequency $\omega_n$.
Thus, the Kramer-Pesch approximation is appropriate near the vortex core
$\rho\ll\xi$ and in low temperature $k_BT\ll|\Delta|$.
In this section, We follow the
method in Ref.~\onlinecite{nagai_JPSJ_2006}.

We start with the Riccati-type differential equations with $\bs{A}=0$:
\begin{align}
	\bs{v}_F\cdot\nabla\gamma+2\omega_n\gamma-\Delta+\Delta^*\gamma^2=0,\label{riccatieq}\\
  \bs{v}_F\cdot\nabla\ut{\gamma}-2\omega_n\ut{\gamma}+s_\mu\Delta^*-s_\mu\Delta\ut{\gamma}^2=0,
\end{align}
where we use the units $\hbar=k_B=1$.
We introduce the coordinate transformation as in Fig.\ref{fig_cotr}:
\begin{align}
  a&=x\cos\phi+y\sin\phi,\\
	b&=-x\sin\phi+y\cos\phi, 
\end{align}
where the angle $\phi$ characterizes the direction of the
momentum: $\hat{\bs{k}} = (\sin\theta\cos\phi, \sin\theta\sin\phi,
\cos\theta)$. The
$b$-axis is perpendicular to
the projected momentum
$\bs{k}_{\parallel}=(\sin\theta\cos\phi,\sin\theta\sin\phi,0)$.
We solve the Riccati-type differential equations by integrating
$\gamma$ or $\ut{\gamma}$ in the direction $\hat{\bs{k}}$.  The
impact parameter $b$ is
constant on the integrating path and corresponds to the distance
between the integrating path and 
vortex core. 
In this coordinate $\bs{r}=(a,b,z)$, the Riccati-type
differential equations are rewritten as
\begin{align}
  & v_F\left\{\sin\theta\frac{\partial}{\partial a}
	+\cos\theta\frac{\partial}{\partial z}\right\}\gamma
	+2\omega_n\gamma-\Delta+\Delta^*\gamma^2=0,\\
  & v_F\left\{\sin\theta\frac{\partial}{\partial a}
	+\cos\theta\frac{\partial}{\partial z}\right\}\ut{\gamma}
	-2\omega_n\ut{\gamma}+s_\mu\Delta^*-s_\mu\Delta\ut{\gamma}^2=0.
\end{align}
In the
analytic calculation, the pair potential is written in more general
form: 
\begin{align}
  \Delta(\hat{\bs{k}})&=h(\rho,z)e^{-i\Phi}\Psi(\hat{\bs{k}}).
\end{align}
where the real function $h$ and the complex function $\Psi$
describe the spatial and momentum dependence of $\Delta$. In
addition, we assume that $\Psi$ satisfies
$s_\theta=\Psi(\theta,\phi)/\Psi(\pi-\theta,\phi)=\pm1$.  In the case
of $s_\theta=1$ and $-1$, the ABSs is absence and present on the
surface perpendicular to the $z$ axis. 
We expand the pair potential as
\begin{align}
  & \Delta(\hat{\bs{k}},\bs{r})=\Delta_0+\Delta_1,\\
  & \Delta_0=h(|a|,z)e^{-i\phi}\mathrm{sign}(a)\Psi(\hat{\bs{k}}),\\
  & \Delta_1=-h(|a|,z)e^{-i\phi}\mathrm{sign}(a)i\frac{b}{a}\Psi(\hat{\bs{k}}),
\end{align}
with respect to $b$.
The solutions \eqref{coh_fun_bulk1} and \eqref{coh_fun_bulk2} in the homogeneous limit are also expanded in terms of 
$b$ and $\omega_n$ as
\begin{widetext}
\begin{align}
  \gamma(\bs{r},\hat{\bs{k}},\omega_n)&=\mathrm{sign}(a\omega_n)\left\{\frac{\Psi(\hat{\bs{k}})e^{-i\phi}}{|\Psi(\hat{\bs{k}})|}-\frac{\Psi(\hat{\bs{k}})e^{-i\phi}}{h(|a|,z)|\Psi(\hat{\bs{k}})|^2}|\omega_n|-i\frac{\Psi(\hat{\bs{k}})e^{-i\phi}}{|\Psi(\hat{\bs{k}})|}\frac{b}{a}\right\}+\cdots,\\
  \ut{\gamma}(\bs{r},\hat{\bs{k}},\omega_n)&=\mathrm{sign}(a\omega_n)\left\{\frac{s_\mu\Psi^*(\hat{\bs{k}})e^{i\phi}}{|\Psi(\hat{\bs{k}})|}-\frac{s_\mu\Psi^*(\hat{\bs{k}})e^{i\phi}}{h(|a|,z)|\Psi(\hat{\bs{k}})|^2}|\omega_n|+i\frac{s_\mu\Psi^*(\hat{\bs{k}})e^{i\phi}}{|\Psi(\hat{\bs{k}})|}\frac{b}{a}\right\}+\cdots.
\end{align}
When we set the cut-off of the integral path as $a_c>0$, the boundary conditions (I) in the Appendix \ref{sec:form} are given as
\begin{align}
	\gamma(a=-a_c,\frac{\pi}{2}<\theta\le\pi,\omega_n>0)&=-\frac{\Psi(\hat{\bs{k}})e^{-i\phi}}{|\Psi(\hat{\bs{k}})|}+\frac{\Psi(\hat{\bs{k}})e^{-i\phi}}{\Delta_0|\Psi(\hat{\bs{k}})|^2}|\omega_n|-i\frac{\Psi(\hat{\bs{k}})e^{-i\phi}}{|\Psi(\hat{\bs{k}})|}\frac{b}{a_c},\label{bc1}\\
  \ut{\gamma}(a=a_c,0\le\theta<\frac{\pi}{2},\omega_n>0)&=\frac{s_\mu\Psi^*(\hat{\bs{k}})e^{i\phi}}{|\Psi(\hat{\bs{k}})|}-\frac{s_\mu\Psi^*(\hat{\bs{k}})e^{i\phi}}{\Delta_0|\Psi(\hat{\bs{k}})|^2}|\omega_n|+i\frac{s_\mu\Psi^*(\hat{\bs{k}})e^{i\phi}}{|\Psi(\hat{\bs{k}})|}\frac{b}{a_c},\label{bc2}
\end{align}
The Riccati-type differential equations of the zeroth order 
of $b$ and $\omega_n$ are written as
\begin{align}
  v_F\left\{\sin\theta\frac{\partial}{\partial a}+\cos\theta\frac{\partial}{\partial z}\right\}\gamma_0-\Delta_0+\Delta^*_0\gamma_0^2=0,\\
  v_F\left\{\sin\theta\frac{\partial}{\partial a}+\cos\theta\frac{\partial}{\partial z}\right\}\ut{\gamma}_0+s_\mu\Delta^*_0-s_\mu\Delta_0\ut{\gamma}^2_0=0.
\end{align}
The solutions of these equations are given by
\begin{align}
\gamma_0(\hat{\bs{k}},\omega_n>0)&
=-\frac{\Psi(\mathrm{max}\{\theta,\pi-\theta\},\phi)}{|\Psi(\hat{\bs{k}})|}e^{-i\phi},\\
\ut{\gamma}_0(\hat{\bs{k}},\omega_n>0)&
=\frac{s_\mu\Psi^*(\mathrm{min}\{\theta,\pi-\theta\},\phi)}{|\Psi(\hat{\bs{k}})|}e^{i\phi},
\end{align}
which satisfies the boundary conditions 
 \eqref{bc}, \eqref{bc1} and \eqref{bc2}.
The first order Riccati-type differential equations are written as
\begin{align}
  v_F\left\{\sin\theta\frac{\partial}{\partial a}+\cos\theta\frac{\partial}{\partial z}\right\}\gamma_1+2\omega_n\gamma_0-\Delta_1+2\Delta^*_0\gamma_0\gamma_1+\Delta_1^*\gamma_0^2=0,\\
  v_F\left\{\sin\theta\frac{\partial}{\partial a}+\cos\theta\frac{\partial}{\partial z}\right\}\ut{\gamma}_1-2\omega_n\ut{\gamma}_0+s_\mu\Delta_1^*-2s_\mu\Delta_0\ut{\gamma}_0\ut{\gamma}_1-s_\mu\Delta_1\ut{\gamma}_0^2=0.
\end{align}
The general solutions of these equations are given as
\begin{align}
	\gamma_1&=\frac{2}{v_F}\exp\left[2|\Psi(\hat{\bs{k}})|F(\sigma,z_0,\theta)\right]\int^\sigma_{\sigma_0}d\sigma^\prime\left[-i\frac{b}{\sigma^\prime\sin\theta}\Delta_0-\omega_n\gamma_0\right]\exp\left[-2|\Psi(\hat{\bs{k}})|F(\sigma^\prime,z_0,\theta)\right]\\&+C_0\exp\left[2|\Psi(\hat{\bs{k}})|F(\sigma,z_0,\theta)\right],\\
  \ut{\gamma}_1&=\frac{2}{v_F}\exp\left[2s_\theta|\Psi(\hat{\bs{k}})|F(\sigma,z_0,\theta)\right]\int^\sigma_{\ut{\sigma}_0}d\sigma^\prime\left[-is_\mu\frac{b}{\sigma^\prime\sin\theta}\Delta^*_0+\omega_n\ut{\gamma}_0\right]\exp\left[-2s_\theta|\Psi(\hat{\bs{k}})|F(\sigma^\prime,z_0,\theta)\right]\\&+\ut{C}_0\exp\left[2s_\theta|\Psi(\hat{\bs{k}})|F(\sigma,z_0,\theta)\right],\\
  F&(\sigma,z_0,\theta)=\int_{-\frac{z_0}{\cos\theta}}^\sigma\mathrm{sign}(\sigma^\prime)\frac{\Psi(\max\{\theta,\pi-\theta\},\phi)}{\Psi(\bs{k})}h(|\sigma^\prime|\sin\theta,\sigma^\prime\cos\theta+z_0)d\sigma^\prime,\\
  a&=\sigma\sin\theta,\\
  z&=\sigma\cos\theta+z_0
\end{align}
where $\sigma_0, C_0,\ut{\sigma}_0$ and $\ut{C}_0$ are integral constants.
In $\pi/2<\theta\le\pi$, we obtain the first order coherence function
$\gamma_1$ from the boundary condition \eqref{bc1} as  
\begin{align}
  \gamma_1&=\frac{2}{v_F}\exp\left[2|\Psi(\hat{\bs{k}})|F(\sigma,z_0,\theta)\right]\int^\sigma_{-\frac{a_c}{\sin\theta}}d\sigma^\prime\left[-i\frac{b}{\sigma^\prime\sin\theta}\Delta_0-\omega_n\gamma_0\right]\exp\left[-2|\Psi(\hat{\bs{k}})|F(\sigma^\prime,z_0,\theta)\right]
        \\&+\left\{\frac{\Psi(\hat{\bs{k}})e^{-i\phi}}{\Delta_\infty|\Psi(\hat{\bs{k}})|^2}|\omega_n|-i\frac{\Psi(\hat{\bs{k}})e^{-i\phi}}{|\Psi(\hat{\bs{k}})|}\frac{b}{a_c}\right\}\exp\left[2|\Psi(\hat{\bs{k}})|\left\{F(\sigma,z_0,\theta)-F\left(-\frac{a_c}{\sin\theta},z_0,\theta\right)\right\}\right].
\end{align}
The other function $\ut{\gamma}_1$ in $0\le\theta<\pi/2$ is given from the boundary condition
\eqref{bc2} as 
\begin{align}
  \ut{\gamma}_1&=-\frac{2}{v_F}\exp\left[2s_\theta|\Psi(\hat{\bs{k}})|F(\sigma,z_0,\theta)\right]\int_\sigma^{\frac{a_c}{\sin\theta}}d\sigma^\prime\left[-is_\mu\frac{b}{\sigma^\prime\sin\theta}\Delta^*_0+\omega_n\ut{\gamma}_0\right]\exp\left[-2s_\theta|\Psi(\hat{\bs{k}})|F(\sigma^\prime,z_0,\theta)\right]
                     \\&-\left\{\frac{s_\mu\Psi^*(\hat{\bs{k}})e^{i\phi}}{\Delta_\infty|\Psi(\hat{\bs{k}})|^2}|\omega_n|-i\frac{s_\mu\Psi^*(\hat{\bs{k}})e^{i\phi}}{|\Psi(\hat{\bs{k}})|}\frac{b}{a_c}\right\}\exp\left[2s_\theta|\Psi(\hat{\bs{k}})|\left\{F(\sigma,z_0,\theta)-F\left(\frac{a_c}{\sin\theta},z_0,\theta\right)\right\}\right].
\end{align}
We respectively set the integral constants as $\sigma_0=-a_c/\sin\theta$ and $\ut{\sigma}_0=a_c/\sin\theta$.
From the boundary condition \eqref{bc}, $\gamma_1$ in $0\le\theta<\pi/2$ and $\ut{\gamma}_1$ in $\pi/2<\theta\le\pi$ are given as
\begin{align}
  \gamma_1&=\frac{2}{v_F}\exp\left[2|\Psi(\hat{\bs{k}})|F(\sigma,z_0,\theta)\right]\int^\sigma_{-\frac{z_0}{\cos\theta}}d\sigma^\prime\left[-i\frac{b}{\sigma^\prime\sin\theta}\Delta_0-\omega_n\gamma_0\right]\exp\left[-2|\Psi(\hat{\bs{k}})|F(\sigma^\prime,z_0,\theta)\right]\\
          &+\frac{2}{v_F}\int^{-\frac{z_0}{\cos\theta}}_{-\frac{a_c}{\sin\theta}}d\sigma^\prime\left[-i\frac{b}{\sigma^\prime\sin\theta}\Delta_0(-z_0,\pi-\theta)-\omega_n\gamma_0\right]\exp\left[2|\Psi(\hat{\bs{k}})|\left\{F(\sigma,z_0,\theta)-F(\sigma^\prime,-z_0,\pi-\theta)\right\}\right]
        \\&+s_\theta\left\{\frac{\Psi(\hat{\bs{k}})e^{-i\phi}}{\Delta_\infty|\Psi(\hat{\bs{k}})|^2}|\omega_n|-i\frac{\Psi(\hat{\bs{k}})e^{-i\phi}}{|\Psi(\hat{\bs{k}})|}\frac{b}{a_c}\right\}\exp\left[2|\Psi(\hat{\bs{k}})|\left\{F(\sigma,z_0,\theta)-F\left(-\frac{a_c}{\sin\theta},-z_0,\pi-\theta\right)\right\}\right],\\
	\ut{\gamma}_1&=-\frac{2}{v_F}\exp\left[2s_\theta|\Psi(\hat{\bs{k}})|F(\sigma,z_0,\theta)\right]\int_\sigma^{-\frac{z_0}{\cos\theta}}d\sigma^\prime\left[-is_\mu\frac{b}{\sigma^\prime\sin\theta}\Delta^*_0+\omega_n\ut{\gamma}_0\right]\exp\left[-2s_\theta|\Psi(\hat{\bs{k}})|F(\sigma^\prime,z_0,\theta)\right]
		     \\&-\frac{2}{v_F}\int_{-\frac{z_0}{\cos\theta}}^{\frac{a_c}{\sin\theta}}d\sigma^\prime\left[-is_\mu\frac{b}{\sigma^\prime\sin\theta}\Delta_0^*(-z_0,\pi-\theta)+\omega_n\ut{\gamma}_0\right]\exp\left[2s_\theta|\Psi(\bs{k})|\left\{F(\sigma,z_0,\theta)-F(\sigma^\prime,-z_0,\pi-\theta)\right\}\right]\\
                       &-s_\theta\left\{\frac{s_\mu\Psi^*(\hat{\bs{k}})e^{i\phi}}{\Delta_\infty|\Psi(\hat{\bs{k}})|^2}|\omega_n|-i\frac{s_\mu\Psi^*(\hat{\bs{k}})e^{i\phi}}{|\Psi(\hat{\bs{k}})|}\frac{b}{a_c}\right\}\exp\left[2s_\theta|\Psi(\hat{\bs{k}})|\left\{F(\sigma,z_0,\theta)-F\left(\frac{a_c}{\sin\theta},-z_0,\pi-\theta\right)\right\}\right].
\end{align}
When we take $z=0,~a_c\rightarrow\infty$, we obtain
\begin{align}
  F\left(\sigma=-\frac{z_0}{\cos\theta},z_0,\theta\right)&=0,\\
  F\left(\sigma=-\infty,z_0,\frac{\pi}{2}<\theta\le\pi\right)&=s_\theta F\left(\sigma=+\infty,z_0,0\le \theta<\frac{\pi}{2}\right)=\infty.
\end{align}
As a result, $\gamma_1$ and $\ut{\gamma}_1$ in $0\le\theta<\pi/2$ is given as
\begin{align}
  \gamma_1&=\frac{2}{v_F}\int^{-\frac{z_0}{\cos\theta}}_{-\infty}d\sigma^\prime\left[-i\frac{b}{\sigma^\prime\sin\theta}\Delta_0(-z_0,\pi-\theta)-\omega_n\gamma_0\right]\exp\left[-2|\Psi(\hat{\bs{k}})|F(\sigma^\prime,-z_0,\pi-\theta)\right],\\
  \ut{\gamma}_1&=-\frac{2}{v_F}\int_{-\frac{z_0}{\cos\theta}}^{\infty}d\sigma^\prime\left[-is_\mu\frac{b}{\sigma^\prime\sin\theta}\Delta^*_0+\omega_n\ut{\gamma}_0\right]\exp\left[-2s_\theta|\Psi(\hat{\bs{k}})|F(\sigma^\prime,z_0,\theta)\right],
\end{align}
and ones in $\pi/2\le\theta<\pi$ is given as
\begin{align}
\gamma_1&=\frac{2}{v_F}\int^{-\frac{z_0}{\cos\theta}}_{-\infty}d\sigma^\prime\left[-i\frac{b}{\sigma^\prime\sin\theta}\Delta_0-\omega_n\gamma_0\right]\exp\left[-2|\Psi(\hat{\bs{k}})|F(\sigma^\prime,z_0,\theta)\right],\\
  \ut{\gamma}_1&=-\frac{2}{v_F}\int_{-\frac{z_0}{\cos\theta}}^{\infty}d\sigma^\prime\left[-is_\mu\frac{b}{\sigma^\prime\sin\theta}\Delta_0^*(-z_0,\pi-\theta)+\omega_n\ut{\gamma}_0\right]\exp\left[-2s_\theta|\Psi(\bs{k})|F(\sigma^\prime,-z_0,\pi-\theta)\right].
\end{align}
We can replace $z$ to $|z|$ due to $z=\sigma\cos\theta+z_0>0$. 
By this substitution, we can rewrite $\gamma_1$ and 
$\ut{\gamma}_1$ in $0\le\theta\le\pi$ as
\begin{align}
  \gamma_1&=\frac{2}{v_F}\int^{-\frac{z_0}{\cos\theta}}_{-\infty}d\sigma^\prime\left[-i\frac{b}{\sigma^\prime\sin\theta}\bar{\Delta}(\sigma^\prime)-\omega_n\gamma_0\right]\exp\left[-2|\Psi(\bs{k})|\bar{F}(\sigma^\prime)\right],\\
\ut{\gamma}_1&=-\frac{2}{v_F}\int_{-\frac{z_0}{\cos\theta}}^{\infty}d\sigma^\prime\left[-is_\mu s_\theta\frac{b}{\sigma^\prime\sin\theta}\bar{\Delta}^*(\sigma^\prime)+\omega_n\ut{\gamma}_0\right]\exp\left[-2|\Psi(\bs{k})|\bar{F}(\sigma^\prime)\right],
\end{align}
where we define $\bar{\Delta}$ and $\bar{F}$ as
\begin{align}
  \bar{\Delta}(\sigma,z_0,\hat{\bs{k}})&=h(|\sigma^\prime|\sin\theta,|\sigma^\prime\cos\theta+z_0|)e^{i\phi}\mathrm{sign}(\sigma)\Psi(\max\{\theta,\pi-\theta\},\phi),\\
  \bar{F}(\sigma,z_0,\hat{\bs{k}})&=\int_{-\frac{z_0}{\cos\theta}}^{\sigma}\mathrm{sign}(\sigma^{\prime})h(|\sigma^{\prime}|\sin\theta,|\sigma^{\prime}\cos\theta+z_0|)d\sigma^{\prime}.
\end{align}

The quasi-classical Green functions are expanded as
\begin{align}
  g&=\frac{1-s_\mu\gamma\ut{\gamma}}{1+s_\mu\gamma\ut{\gamma}}
  \simeq\frac{1-s_\mu\left(\gamma_0\ut{\gamma}_0+\gamma_1\ut{\gamma}_0+\gamma_0\ut{\gamma}_1\right)}{1+s_\mu\left(\gamma_0\ut{\gamma}_0+\gamma_1\ut{\gamma}_0+\gamma_0\ut{\gamma}_1\right)},\\
  f&=\frac{2\gamma}{1+s_\mu\gamma\ut{\gamma}}
  \simeq\frac{2\gamma_0+2\gamma_1}{1+s_\mu\left(\gamma_0\ut{\gamma}_0+\gamma_1\ut{\gamma}_0+\gamma_0\ut{\gamma}_1\right)}.
\end{align}
The relation $\gamma_0\ut{\gamma}_0=-s_\mu
s_\theta$ is important.
In the case of $s_\theta=1$, zeroth order Green functions diverge,
whereas ones do not in the case of $s_\theta=-1$.  
We consider the case of $s_\theta=-1$. In this case, the first order
Green functions are approximated as
\begin{align}
	g\simeq-\frac{s_\mu W}{2+s_\mu W},
	\hspace{9mm}
	f\simeq\frac{2\gamma_0+2\gamma_1}{2+s_\mu W},\label{fexpand}
\end{align}
where we define
\begin{align}
	W(b,z_0,\bs{k})&=\gamma_1\ut{\gamma}_0+\gamma_0\ut{\gamma}_1
		     =\frac{2}{v_F}\int^{\infty}_{-\infty}d\sigma^\prime\left[-i\frac{b}{\sigma^\prime\sin\theta}\bar{\Delta}(\sigma^\prime)\ut{\gamma}_0-s_\mu\omega_n\right]\exp\left[-2|\Psi(\bs{k})|\bar{F}(\sigma^\prime)\right].
\end{align}
In the cylindrical coordinate $\bs{r}=(\rho,\Phi,z=0)$, $W$ is rewritten as
\begin{align}
  W(\rho,\Phi,z=0,\bs{k})&=-2s_\mu(i\rho CD+\omega_nC),
\end{align}
where we use $b=\rho\sin(\phi-\Phi)$ and define $D$ and $C$ as
\begin{align}
  D(\rho,\Phi,\bs{k})&=\frac{2}{C(\rho,\Phi,\bs{k})}\int^{\infty}_{-\infty}d\sigma^\prime\frac{\sin(\phi-\Phi)}{\sigma^\prime\sin\theta}h\left(|\sigma|\sin\theta,\left|\sigma\cos\theta-\rho\frac{\cos(\phi-\Phi)}{\tan\theta}\right|\right)\exp\left[-2|\Psi(\bs{k})|\bar{F}(\rho,\Phi,\sigma^\prime,\bs{k})\right]\\
		  &=\frac{\sin(\phi-\Phi)}{\sin\theta}\tilde{D}(\rho,\Phi,\bs{k}),\\
  C(\rho,\Phi,\bs{k})&=\frac{2}{v_F}\int^{\infty}_{-\infty}d\sigma^\prime\exp\left[-2|\Psi(\bs{k})|\bar{F}(\rho,\Phi,\sigma^\prime,\bs{k})\right].
\end{align}
It is noted that $\tilde{D},~C>0$ due to $h>0$.
The azimuthal component of the current density is written as
\begin{align}
	j_\Phi(\bs{r})&=-4|e|N_0\pi T\sum_{n=0}^{\infty}\int\frac{d\Omega}{4\pi}v_F\sin\theta\sin(\phi-\Phi) \mathrm{Im}g(\hat{\bs{k}},\bs{r},\omega_n). 
\end{align}
In order to calculate the current density, we perform the sum over the Matsubara frequencies as 
\begin{align}
	-T\sin\theta\sin(\phi-\Phi)\sum_{n=1}^\infty\mathrm{Im}g(\rho,\Phi,\bs{k},i\omega_n)&=\sum_{n=1}^\infty\frac{T\sin\theta\sin(\phi-\Phi)\rho C^{-1}D}{(i\omega_n-\rho D-iC^{-1})(i\omega_n+\rho D-iC^{-1})}\\
				      &=\frac{C^{-1}}{4}\sin\theta\sin(\phi-\Phi)
	\mathrm{Re}\left[\tanh\frac{1}{2T}\left(-\rho D+iC^{-1}\right)\right]
				    \\&\quad+\frac{1}{\pi}\int_0^\infty\frac{\sin^2(\phi-\Phi)\rho C^{-1}\tilde{D}z}{\{(z+\rho D)^2+C^{-2}\}\{(z-\rho D)^2+C^{-2}\}}\tanh\frac{z}{2T}dz.
\end{align}
The second term is bounded from above as
\begin{align}
				      &\quad\frac{1}{\pi}\int_0^\infty\frac{\sin^2(\phi-\Phi)\rho C^{-1}\tilde{D}z}{\{(z+\rho D)^2+C^{-2}\}\{(z-\rho D)^2+C^{-2}\}}\tanh\frac{z}{2T}dz\\
				      &\le\frac{1}{\pi}\int_0^\infty\frac{\sin^2(\phi-\Phi)\rho C^{-1}\tilde{D}z}{\{(z+\rho D)^2+C^{-2}\}\{(z-\rho D)^2+C^{-2}\}}dz
				    \\&=\frac{1}{4\pi}\sin\theta\sin(\phi-\Phi)\int_{-\rho D}^{\rho D}\frac{C^{-1}}{z^2+C^{-2}}dz=2\sin\theta\sin(\phi-\Phi)\tan^{-1}\left(\frac{\rho D}{C^{-1}}\right).
\end{align}
From this inequality, we obtain
\begin{align}
	&\quad-T\sin\theta\sin(\phi-\Phi)\sum_{n=1}^\infty\mathrm{Im}g(\rho,\Phi,\bs{k},i\omega_n)
	\\&\le\frac{C^{-1}}{4}\sin\theta\sin(\phi-\Phi)\mathrm{Re}\left[\tanh\frac{1}{2T}\left(-\rho D+iC^{-1}\right)\right]+2\sin\theta\sin(\phi-\Phi)\tan^{-1}\left(\frac{\rho D}{C^{-1}}\right)
	\\&\simeq\sin^2(\phi-\Phi)\left[-\frac{C^{-1}}{8T}+\frac{2}{C^{-1}}\right]\rho\tilde{D}+O(\rho^2).
\end{align}
Therefore, the vortex current at $\rho\ll\xi$ and
$T\ll\Delta$ is bounded from above as follows:
\begin{align}
	j_\Phi(\rho)&\le4|e|N_0\pi v_F\int\frac{d\Omega}{4\pi}\sin^2(\phi-\Phi)\left[-\frac{C^{-1}}{8T}+\frac{2}{C^{-1}}\right]\rho\tilde{D}.\label{jseq-1}
\end{align}
On the other hand, in the case of $s_\theta=1$ or deep inside the SC, the current density is given by
\begin{align}
	j_\Phi(\rho)&=4|e|N_0\pi v_F\int\frac{d\Omega}{4\pi}\sin\theta\sin(\phi-\Phi)\frac{C^{-1}}{2}\tanh\left(\frac{\rho D}{2T}\right)\\
	      &\simeq4|e|N_0\pi v_F\int\frac{d\Omega}{4\pi}\sin^2(\phi-\Phi)\frac{C^{-1}}{4T}\rho\tilde{D}+O(\rho^2),\label{jseq1}
\end{align}
\end{widetext}
where $N_0$ is the density of state in the normal state.
These equations \eqref{jseq-1} and \eqref{jseq1}
show that the vortex current is reversed
near the surface when $s_\theta=-1$.

The LDOS is written as
\begin{align}
	\frac{N(E,\rho)}{N_0}&=-\lim_{\delta\rightarrow0}\int\frac{d\Omega}{4\pi}\mathrm{Re}g(\bs{k},\bs{r},E+i\delta).
\end{align}
From Eq.~\eqref{fexpand}, the LDOS is calculated as
\begin{align}
	\frac{N(E,\rho)}{N_0}&=-\lim_{\delta\rightarrow0}\int\frac{d\Omega}{4\pi}\mathrm{Re}\frac{\rho D+E+i\delta}{iC^{-1}-\rho D-E-i\delta}
			   \\&=\int\frac{d\Omega}{4\pi}\frac{(\rho D+E)^2}{(\rho D+E)^2+C^{-2}}\le1.
\end{align}
Therefore, the LDOS is
suppressed significantly at $\rho\ll\xi$ and $E\ll|\Delta|$ and equals zero especially at $\rho=0$ and $E=0$.
These results are consistent with numerical results for the
$p_z$-wave SC in the main texts.

\bibliographystyle{apsrev4-1}
\bibliography{Yoshida_no_bib}

\begin{thebibliography}{58}%
\makeatletter
\providecommand \@ifxundefined [1]{%
 \@ifx{#1\undefined}
}%
\providecommand \@ifnum [1]{%
 \ifnum #1\expandafter \@firstoftwo
 \else \expandafter \@secondoftwo
 \fi
}%
\providecommand \@ifx [1]{%
 \ifx #1\expandafter \@firstoftwo
 \else \expandafter \@secondoftwo
 \fi
}%
\providecommand \natexlab [1]{#1}%
\providecommand \enquote  [1]{``#1''}%
\providecommand \bibnamefont  [1]{#1}%
\providecommand \bibfnamefont [1]{#1}%
\providecommand \citenamefont [1]{#1}%
\providecommand \href@noop [0]{\@secondoftwo}%
\providecommand \href [0]{\begingroup \@sanitize@url \@href}%
\providecommand \@href[1]{\@@startlink{#1}\@@href}%
\providecommand \@@href[1]{\endgroup#1\@@endlink}%
\providecommand \@sanitize@url [0]{\catcode `\\12\catcode `\$12\catcode
  `\&12\catcode `\#12\catcode `\^12\catcode `\_12\catcode `\%12\relax}%
\providecommand \@@startlink[1]{}%
\providecommand \@@endlink[0]{}%
\providecommand \url  [0]{\begingroup\@sanitize@url \@url }%
\providecommand \@url [1]{\endgroup\@href {#1}{\urlprefix }}%
\providecommand \urlprefix  [0]{URL }%
\providecommand \Eprint [0]{\href }%
\providecommand \doibase [0]{http://dx.doi.org/}%
\providecommand \selectlanguage [0]{\@gobble}%
\providecommand \bibinfo  [0]{\@secondoftwo}%
\providecommand \bibfield  [0]{\@secondoftwo}%
\providecommand \translation [1]{[#1]}%
\providecommand \BibitemOpen [0]{}%
\providecommand \bibitemStop [0]{}%
\providecommand \bibitemNoStop [0]{.\EOS\space}%
\providecommand \EOS [0]{\spacefactor3000\relax}%
\providecommand \BibitemShut  [1]{\csname bibitem#1\endcsname}%
\let\auto@bib@innerbib\@empty
\bibitem [{\citenamefont {Di~Bernardo}\ \emph {et~al.}(2015)\citenamefont
  {Di~Bernardo}, \citenamefont {Salman}, \citenamefont {Wang}, \citenamefont
  {Amado}, \citenamefont {Egilmez}, \citenamefont {Flokstra}, \citenamefont
  {Suter}, \citenamefont {Lee}, \citenamefont {Zhao}, \citenamefont {Prokscha},
  \citenamefont {Morenzoni}, \citenamefont {Blamire}, \citenamefont {Linder},\
  and\ \citenamefont {Robinson}}]{Bernardo_PRX_2015}%
  \BibitemOpen
  \bibfield  {author} {\bibinfo {author} {\bibfnamefont {A.}~\bibnamefont
  {Di~Bernardo}}, \bibinfo {author} {\bibfnamefont {Z.}~\bibnamefont {Salman}},
  \bibinfo {author} {\bibfnamefont {X.~L.}\ \bibnamefont {Wang}}, \bibinfo
  {author} {\bibfnamefont {M.}~\bibnamefont {Amado}}, \bibinfo {author}
  {\bibfnamefont {M.}~\bibnamefont {Egilmez}}, \bibinfo {author} {\bibfnamefont
  {M.~G.}\ \bibnamefont {Flokstra}}, \bibinfo {author} {\bibfnamefont
  {A.}~\bibnamefont {Suter}}, \bibinfo {author} {\bibfnamefont {S.~L.}\
  \bibnamefont {Lee}}, \bibinfo {author} {\bibfnamefont {J.~H.}\ \bibnamefont
  {Zhao}}, \bibinfo {author} {\bibfnamefont {T.}~\bibnamefont {Prokscha}},
  \bibinfo {author} {\bibfnamefont {E.}~\bibnamefont {Morenzoni}}, \bibinfo
  {author} {\bibfnamefont {M.~G.}\ \bibnamefont {Blamire}}, \bibinfo {author}
  {\bibfnamefont {J.}~\bibnamefont {Linder}}, \ and\ \bibinfo {author}
  {\bibfnamefont {J.~W.~A.}\ \bibnamefont {Robinson}},\ }\href {\doibase
  10.1103/PhysRevX.5.041021} {\bibfield  {journal} {\bibinfo  {journal} {Phys.
  Rev. X}\ }\textbf {\bibinfo {volume} {5}},\ \bibinfo {pages} {041021}
  (\bibinfo {year} {2015})}\BibitemShut {NoStop}%
\bibitem [{\citenamefont {Krieger}\ \emph {et~al.}(2020)\citenamefont
  {Krieger}, \citenamefont {Pertsova}, \citenamefont {Giblin}, \citenamefont
  {D\"obeli}, \citenamefont {Prokscha}, \citenamefont {Schneider},
  \citenamefont {Suter}, \citenamefont {Hesjedal}, \citenamefont {Balatsky},\
  and\ \citenamefont {Salman}}]{Krieger_PRL_2020}%
  \BibitemOpen
  \bibfield  {author} {\bibinfo {author} {\bibfnamefont {J.~A.}\ \bibnamefont
  {Krieger}}, \bibinfo {author} {\bibfnamefont {A.}~\bibnamefont {Pertsova}},
  \bibinfo {author} {\bibfnamefont {S.~R.}\ \bibnamefont {Giblin}}, \bibinfo
  {author} {\bibfnamefont {M.}~\bibnamefont {D\"obeli}}, \bibinfo {author}
  {\bibfnamefont {T.}~\bibnamefont {Prokscha}}, \bibinfo {author}
  {\bibfnamefont {C.~W.}\ \bibnamefont {Schneider}}, \bibinfo {author}
  {\bibfnamefont {A.}~\bibnamefont {Suter}}, \bibinfo {author} {\bibfnamefont
  {T.}~\bibnamefont {Hesjedal}}, \bibinfo {author} {\bibfnamefont {A.~V.}\
  \bibnamefont {Balatsky}}, \ and\ \bibinfo {author} {\bibfnamefont
  {Z.}~\bibnamefont {Salman}},\ }\href {\doibase
  10.1103/PhysRevLett.125.026802} {\bibfield  {journal} {\bibinfo  {journal}
  {Phys. Rev. Lett.}\ }\textbf {\bibinfo {volume} {125}},\ \bibinfo {pages}
  {026802} (\bibinfo {year} {2020})}\BibitemShut {NoStop}%
\bibitem [{\citenamefont {Berezinskii}()}]{Berezinskii}%
  \BibitemOpen
  \bibfield  {author} {\bibinfo {author} {\bibfnamefont {V.~L.}\ \bibnamefont
  {Berezinskii}},\ }\href@noop {} {\bibinfo  {journal} {Pis¿ma Zh. Eksp. Teor.
  Fiz. \textbf{20}, 628 (1974) [JETP Lett.20, 287 (1974)]}\ }\BibitemShut
  {NoStop}%
\bibitem [{\citenamefont {Balatsky}\ and\ \citenamefont
  {Abrahams}(1992)}]{balatsky_1992_PRB}%
  \BibitemOpen
\bibfield  {journal} {  }\bibfield  {author} {\bibinfo {author} {\bibfnamefont
  {A.}~\bibnamefont {Balatsky}}\ and\ \bibinfo {author} {\bibfnamefont
  {E.}~\bibnamefont {Abrahams}},\ }\href {\doibase 10.1103/PhysRevB.45.13125}
  {\bibfield  {journal} {\bibinfo  {journal} {Phys. Rev. B}\ }\textbf {\bibinfo
  {volume} {45}},\ \bibinfo {pages} {13125} (\bibinfo {year}
  {1992})}\BibitemShut {NoStop}%
\bibitem [{\citenamefont {Linder}\ and\ \citenamefont
  {Balatsky}(2019)}]{Linder_2019_RMP}%
  \BibitemOpen
  \bibfield  {author} {\bibinfo {author} {\bibfnamefont {J.}~\bibnamefont
  {Linder}}\ and\ \bibinfo {author} {\bibfnamefont {A.~V.}\ \bibnamefont
  {Balatsky}},\ }\href {\doibase 10.1103/RevModPhys.91.045005} {\bibfield
  {journal} {\bibinfo  {journal} {Rev. Mod. Phys.}\ }\textbf {\bibinfo {volume}
  {91}},\ \bibinfo {pages} {045005} (\bibinfo {year} {2019})}\BibitemShut
  {NoStop}%
\bibitem [{\citenamefont {Asano}\ \emph {et~al.}(2011)\citenamefont {Asano},
  \citenamefont {Golubov}, \citenamefont {Fominov},\ and\ \citenamefont
  {Tanaka}}]{Asano_PRL_2011}%
  \BibitemOpen
  \bibfield  {author} {\bibinfo {author} {\bibfnamefont {Y.}~\bibnamefont
  {Asano}}, \bibinfo {author} {\bibfnamefont {A.~A.}\ \bibnamefont {Golubov}},
  \bibinfo {author} {\bibfnamefont {Y.~V.}\ \bibnamefont {Fominov}}, \ and\
  \bibinfo {author} {\bibfnamefont {Y.}~\bibnamefont {Tanaka}},\ }\href
  {\doibase 10.1103/PhysRevLett.107.087001} {\bibfield  {journal} {\bibinfo
  {journal} {Phys. Rev. Lett.}\ }\textbf {\bibinfo {volume} {107}},\ \bibinfo
  {pages} {087001} (\bibinfo {year} {2011})}\BibitemShut {NoStop}%
\bibitem [{\citenamefont {Yokoyama}\ \emph {et~al.}(2011)\citenamefont
  {Yokoyama}, \citenamefont {Tanaka},\ and\ \citenamefont
  {Nagaosa}}]{Yokoyama_PRL_2011}%
  \BibitemOpen
  \bibfield  {author} {\bibinfo {author} {\bibfnamefont {T.}~\bibnamefont
  {Yokoyama}}, \bibinfo {author} {\bibfnamefont {Y.}~\bibnamefont {Tanaka}}, \
  and\ \bibinfo {author} {\bibfnamefont {N.}~\bibnamefont {Nagaosa}},\ }\href
  {\doibase 10.1103/PhysRevLett.106.246601} {\bibfield  {journal} {\bibinfo
  {journal} {Phys. Rev. Lett.}\ }\textbf {\bibinfo {volume} {106}},\ \bibinfo
  {pages} {246601} (\bibinfo {year} {2011})}\BibitemShut {NoStop}%
\bibitem [{\citenamefont {Higashitani}\ \emph {et~al.}(2013)\citenamefont
  {Higashitani}, \citenamefont {Takeuchi}, \citenamefont {Matsuo},
  \citenamefont {Nagato},\ and\ \citenamefont {Nagai}}]{Higashitani_PRL_2013}%
  \BibitemOpen
  \bibfield  {author} {\bibinfo {author} {\bibfnamefont {S.}~\bibnamefont
  {Higashitani}}, \bibinfo {author} {\bibfnamefont {H.}~\bibnamefont
  {Takeuchi}}, \bibinfo {author} {\bibfnamefont {S.}~\bibnamefont {Matsuo}},
  \bibinfo {author} {\bibfnamefont {Y.}~\bibnamefont {Nagato}}, \ and\ \bibinfo
  {author} {\bibfnamefont {K.}~\bibnamefont {Nagai}},\ }\href {\doibase
  10.1103/PhysRevLett.110.175301} {\bibfield  {journal} {\bibinfo  {journal}
  {Phys. Rev. Lett.}\ }\textbf {\bibinfo {volume} {110}},\ \bibinfo {pages}
  {175301} (\bibinfo {year} {2013})}\BibitemShut {NoStop}%
\bibitem [{\citenamefont
  {Higashitani}(2014{\natexlab{a}})}]{Higashitani_PRB_2014}%
  \BibitemOpen
  \bibfield  {author} {\bibinfo {author} {\bibfnamefont {S.}~\bibnamefont
  {Higashitani}},\ }\href {\doibase 10.1103/PhysRevB.89.184505} {\bibfield
  {journal} {\bibinfo  {journal} {Phys. Rev. B}\ }\textbf {\bibinfo {volume}
  {89}},\ \bibinfo {pages} {184505} (\bibinfo {year}
  {2014}{\natexlab{a}})}\BibitemShut {NoStop}%
\bibitem [{\citenamefont {Suzuki}\ and\ \citenamefont
  {Asano}(2014)}]{Suzuki_PRB_2014}%
  \BibitemOpen
  \bibfield  {author} {\bibinfo {author} {\bibfnamefont {S.-I.}\ \bibnamefont
  {Suzuki}}\ and\ \bibinfo {author} {\bibfnamefont {Y.}~\bibnamefont {Asano}},\
  }\href {\doibase 10.1103/PhysRevB.89.184508} {\bibfield  {journal} {\bibinfo
  {journal} {Phys. Rev. B}\ }\textbf {\bibinfo {volume} {89}},\ \bibinfo
  {pages} {184508} (\bibinfo {year} {2014})}\BibitemShut {NoStop}%
\bibitem [{\citenamefont {Suzuki}\ and\ \citenamefont
  {Asano}(2015)}]{Suzuki_PRB_2015}%
  \BibitemOpen
  \bibfield  {author} {\bibinfo {author} {\bibfnamefont {S.-I.}\ \bibnamefont
  {Suzuki}}\ and\ \bibinfo {author} {\bibfnamefont {Y.}~\bibnamefont {Asano}},\
  }\href {\doibase 10.1103/PhysRevB.91.214510} {\bibfield  {journal} {\bibinfo
  {journal} {Phys. Rev. B}\ }\textbf {\bibinfo {volume} {91}},\ \bibinfo
  {pages} {214510} (\bibinfo {year} {2015})}\BibitemShut {NoStop}%
\bibitem [{\citenamefont {Asano}\ and\ \citenamefont
  {Sasaki}(2015)}]{Asano_PRB_2015}%
  \BibitemOpen
  \bibfield  {author} {\bibinfo {author} {\bibfnamefont {Y.}~\bibnamefont
  {Asano}}\ and\ \bibinfo {author} {\bibfnamefont {A.}~\bibnamefont {Sasaki}},\
  }\href {\doibase 10.1103/PhysRevB.92.224508} {\bibfield  {journal} {\bibinfo
  {journal} {Phys. Rev. B}\ }\textbf {\bibinfo {volume} {92}},\ \bibinfo
  {pages} {224508} (\bibinfo {year} {2015})}\BibitemShut {NoStop}%
\bibitem [{\citenamefont {Tanaka}\ and\ \citenamefont
  {Kashiwaya}(1995)}]{tanaka_PRL_1995}%
  \BibitemOpen
  \bibfield  {author} {\bibinfo {author} {\bibfnamefont {Y.}~\bibnamefont
  {Tanaka}}\ and\ \bibinfo {author} {\bibfnamefont {S.}~\bibnamefont
  {Kashiwaya}},\ }\href {\doibase 10.1103/PhysRevLett.74.3451} {\bibfield
  {journal} {\bibinfo  {journal} {Phys. Rev. Lett.}\ }\textbf {\bibinfo
  {volume} {74}},\ \bibinfo {pages} {3451} (\bibinfo {year}
  {1995})}\BibitemShut {NoStop}%
\bibitem [{\citenamefont {Tanaka}\ \emph {et~al.}(2005)\citenamefont {Tanaka},
  \citenamefont {Asano}, \citenamefont {Golubov},\ and\ \citenamefont
  {Kashiwaya}}]{tanaka_PRB_2005}%
  \BibitemOpen
  \bibfield  {author} {\bibinfo {author} {\bibfnamefont {Y.}~\bibnamefont
  {Tanaka}}, \bibinfo {author} {\bibfnamefont {Y.}~\bibnamefont {Asano}},
  \bibinfo {author} {\bibfnamefont {A.~A.}\ \bibnamefont {Golubov}}, \ and\
  \bibinfo {author} {\bibfnamefont {S.}~\bibnamefont {Kashiwaya}},\ }\href
  {\doibase 10.1103/PhysRevB.72.140503} {\bibfield  {journal} {\bibinfo
  {journal} {Phys. Rev. B}\ }\textbf {\bibinfo {volume} {72}},\ \bibinfo
  {pages} {140503} (\bibinfo {year} {2005})}\BibitemShut {NoStop}%
\bibitem [{\citenamefont {Tanaka}\ and\ \citenamefont
  {Golubov}(2007)}]{tanaka_PRL_2007}%
  \BibitemOpen
  \bibfield  {author} {\bibinfo {author} {\bibfnamefont {Y.}~\bibnamefont
  {Tanaka}}\ and\ \bibinfo {author} {\bibfnamefont {A.~A.}\ \bibnamefont
  {Golubov}},\ }\href {\doibase 10.1103/PhysRevLett.98.037003} {\bibfield
  {journal} {\bibinfo  {journal} {Phys. Rev. Lett.}\ }\textbf {\bibinfo
  {volume} {98}},\ \bibinfo {pages} {037003} (\bibinfo {year}
  {2007})}\BibitemShut {NoStop}%
\bibitem [{\citenamefont {Eschrig}\ and\ \citenamefont
  {L{\"o}fwander}(2008)}]{Eschrig_Nat_2007}%
  \BibitemOpen
  \bibfield  {author} {\bibinfo {author} {\bibfnamefont {M.}~\bibnamefont
  {Eschrig}}\ and\ \bibinfo {author} {\bibfnamefont {T.}~\bibnamefont
  {L{\"o}fwander}},\ }\href@noop {} {\bibfield  {journal} {\bibinfo  {journal}
  {Nature Physics}\ }\textbf {\bibinfo {volume} {4}},\ \bibinfo {pages} {138}
  (\bibinfo {year} {2008})}\BibitemShut {NoStop}%
\bibitem [{\citenamefont {Suzuki}\ and\ \citenamefont
  {Asano}(2016)}]{Suzuki_PRB_2016}%
  \BibitemOpen
  \bibfield  {author} {\bibinfo {author} {\bibfnamefont {S.-I.}\ \bibnamefont
  {Suzuki}}\ and\ \bibinfo {author} {\bibfnamefont {Y.}~\bibnamefont {Asano}},\
  }\href {\doibase 10.1103/PhysRevB.94.155302} {\bibfield  {journal} {\bibinfo
  {journal} {Phys. Rev. B}\ }\textbf {\bibinfo {volume} {94}},\ \bibinfo
  {pages} {155302} (\bibinfo {year} {2016})}\BibitemShut {NoStop}%
\bibitem [{\citenamefont {Asano}\ \emph {et~al.}(2007)\citenamefont {Asano},
  \citenamefont {Tanaka},\ and\ \citenamefont {Golubov}}]{Asano_PRL_2007}%
  \BibitemOpen
  \bibfield  {author} {\bibinfo {author} {\bibfnamefont {Y.}~\bibnamefont
  {Asano}}, \bibinfo {author} {\bibfnamefont {Y.}~\bibnamefont {Tanaka}}, \
  and\ \bibinfo {author} {\bibfnamefont {A.~A.}\ \bibnamefont {Golubov}},\
  }\href {\doibase 10.1103/PhysRevLett.98.107002} {\bibfield  {journal}
  {\bibinfo  {journal} {Phys. Rev. Lett.}\ }\textbf {\bibinfo {volume} {98}},\
  \bibinfo {pages} {107002} (\bibinfo {year} {2007})}\BibitemShut {NoStop}%
\bibitem [{\citenamefont {Tanaka}\ \emph {et~al.}(2012)\citenamefont {Tanaka},
  \citenamefont {Sato},\ and\ \citenamefont {Nagaosa}}]{tanaka_JPSJ_2012}%
  \BibitemOpen
  \bibfield  {author} {\bibinfo {author} {\bibfnamefont {Y.}~\bibnamefont
  {Tanaka}}, \bibinfo {author} {\bibfnamefont {M.}~\bibnamefont {Sato}}, \ and\
  \bibinfo {author} {\bibfnamefont {N.}~\bibnamefont {Nagaosa}},\ }\href
  {\doibase 10.1143/JPSJ.81.011013} {\bibfield  {journal} {\bibinfo  {journal}
  {J. Phys. Soc. Jpn.}\ }\textbf {\bibinfo {volume} {81}},\ \bibinfo {pages}
  {011013} (\bibinfo {year} {2012})}\BibitemShut {NoStop}%
\bibitem [{\citenamefont {Suzuki}\ \emph {et~al.}(2021)\citenamefont {Suzuki},
  \citenamefont {Hirai}, \citenamefont {Eschrig},\ and\ \citenamefont
  {Tanaka}}]{Suzuki_PRR_2021}%
  \BibitemOpen
  \bibfield  {author} {\bibinfo {author} {\bibfnamefont {S.-I.}\ \bibnamefont
  {Suzuki}}, \bibinfo {author} {\bibfnamefont {T.}~\bibnamefont {Hirai}},
  \bibinfo {author} {\bibfnamefont {M.}~\bibnamefont {Eschrig}}, \ and\
  \bibinfo {author} {\bibfnamefont {Y.}~\bibnamefont {Tanaka}},\ }\href
  {\doibase 10.1103/PhysRevResearch.3.043148} {\bibfield  {journal} {\bibinfo
  {journal} {Phys. Rev. Research}\ }\textbf {\bibinfo {volume} {3}},\ \bibinfo
  {pages} {043148} (\bibinfo {year} {2021})}\BibitemShut {NoStop}%
\bibitem [{\citenamefont {Rogers}\ \emph {et~al.}(2021)\citenamefont {Rogers},
  \citenamefont {Walton}, \citenamefont {Flokstra}, \citenamefont {Ma'Mari},
  \citenamefont {Stewart}, \citenamefont {Lee}, \citenamefont {Prokscha},
  \citenamefont {Caruana}, \citenamefont {Kinane}, \citenamefont {Langridge}, ,
  \citenamefont {Bradshaw}, \citenamefont {Moorsom}, \citenamefont {Ali},
  \citenamefont {Burnell}, \citenamefont {Hickey},\ and\ \citenamefont
  {Cespedes}}]{rogers2021spin}%
  \BibitemOpen
  \bibfield  {author} {\bibinfo {author} {\bibfnamefont {M.}~\bibnamefont
  {Rogers}}, \bibinfo {author} {\bibfnamefont {A.}~\bibnamefont {Walton}},
  \bibinfo {author} {\bibfnamefont {M.~G.}\ \bibnamefont {Flokstra}}, \bibinfo
  {author} {\bibfnamefont {A.}~\bibnamefont {Ma'Mari}}, \bibinfo {author}
  {\bibfnamefont {R.}~\bibnamefont {Stewart}}, \bibinfo {author} {\bibfnamefont
  {S.~L.}\ \bibnamefont {Lee}}, \bibinfo {author} {\bibfnamefont
  {T.}~\bibnamefont {Prokscha}}, \bibinfo {author} {\bibfnamefont {A.~J.}\
  \bibnamefont {Caruana}}, \bibinfo {author} {\bibfnamefont {C.~J.}\
  \bibnamefont {Kinane}}, \bibinfo {author} {\bibfnamefont {S.}~\bibnamefont
  {Langridge}}, , \bibinfo {author} {\bibfnamefont {H.}~\bibnamefont
  {Bradshaw}}, \bibinfo {author} {\bibfnamefont {T.}~\bibnamefont {Moorsom}},
  \bibinfo {author} {\bibfnamefont {M.}~\bibnamefont {Ali}}, \bibinfo {author}
  {\bibfnamefont {G.}~\bibnamefont {Burnell}}, \bibinfo {author} {\bibfnamefont
  {B.~J.}\ \bibnamefont {Hickey}}, \ and\ \bibinfo {author} {\bibfnamefont
  {O.}~\bibnamefont {Cespedes}},\ }\href {\doibase
  https://doi.org/10.1038/s42005-021-00567-7} {\bibfield  {journal} {\bibinfo
  {journal} {Communications Physics}\ }\textbf {\bibinfo {volume} {4}},\
  \bibinfo {pages} {1} (\bibinfo {year} {2021})}\BibitemShut {NoStop}%
\bibitem [{\citenamefont {Yokoyama}\ \emph
  {et~al.}(2008{\natexlab{a}})\citenamefont {Yokoyama}, \citenamefont
  {Tanaka},\ and\ \citenamefont {Golubov}}]{yokoyama_PRB_2008}%
  \BibitemOpen
  \bibfield  {author} {\bibinfo {author} {\bibfnamefont {T.}~\bibnamefont
  {Yokoyama}}, \bibinfo {author} {\bibfnamefont {Y.}~\bibnamefont {Tanaka}}, \
  and\ \bibinfo {author} {\bibfnamefont {A.~A.}\ \bibnamefont {Golubov}},\
  }\href {\doibase 10.1103/PhysRevB.78.012508} {\bibfield  {journal} {\bibinfo
  {journal} {Phys. Rev. B}\ }\textbf {\bibinfo {volume} {78}},\ \bibinfo
  {pages} {012508} (\bibinfo {year} {2008}{\natexlab{a}})}\BibitemShut
  {NoStop}%
\bibitem [{\citenamefont {Tanuma}\ \emph {et~al.}(2009)\citenamefont {Tanuma},
  \citenamefont {Hayashi}, \citenamefont {Tanaka},\ and\ \citenamefont
  {Golubov}}]{tanuma_PRL_2009}%
  \BibitemOpen
  \bibfield  {author} {\bibinfo {author} {\bibfnamefont {Y.}~\bibnamefont
  {Tanuma}}, \bibinfo {author} {\bibfnamefont {N.}~\bibnamefont {Hayashi}},
  \bibinfo {author} {\bibfnamefont {Y.}~\bibnamefont {Tanaka}}, \ and\ \bibinfo
  {author} {\bibfnamefont {A.~A.}\ \bibnamefont {Golubov}},\ }\href {\doibase
  10.1103/PhysRevLett.102.117003} {\bibfield  {journal} {\bibinfo  {journal}
  {Phys. Rev. Lett.}\ }\textbf {\bibinfo {volume} {102}},\ \bibinfo {pages}
  {117003} (\bibinfo {year} {2009})}\BibitemShut {NoStop}%
\bibitem [{\citenamefont {Tanaka}\ \emph {et~al.}(2007)\citenamefont {Tanaka},
  \citenamefont {Tanuma},\ and\ \citenamefont {Golubov}}]{tanaka_PRB_2007}%
  \BibitemOpen
  \bibfield  {author} {\bibinfo {author} {\bibfnamefont {Y.}~\bibnamefont
  {Tanaka}}, \bibinfo {author} {\bibfnamefont {Y.}~\bibnamefont {Tanuma}}, \
  and\ \bibinfo {author} {\bibfnamefont {A.~A.}\ \bibnamefont {Golubov}},\
  }\href {\doibase 10.1103/PhysRevB.76.054522} {\bibfield  {journal} {\bibinfo
  {journal} {Phys. Rev. B}\ }\textbf {\bibinfo {volume} {76}},\ \bibinfo
  {pages} {054522} (\bibinfo {year} {2007})}\BibitemShut {NoStop}%
\bibitem [{\citenamefont {Eschrig}\ \emph {et~al.}(2007)\citenamefont
  {Eschrig}, \citenamefont {L{\"o}fwander}, \citenamefont {Champel},
  \citenamefont {Cuevas}, \citenamefont {Kopu},\ and\ \citenamefont
  {Sch{\"o}n}}]{eschrig_JLTP_2007}%
  \BibitemOpen
  \bibfield  {author} {\bibinfo {author} {\bibfnamefont {M.}~\bibnamefont
  {Eschrig}}, \bibinfo {author} {\bibfnamefont {T.}~\bibnamefont
  {L{\"o}fwander}}, \bibinfo {author} {\bibfnamefont {T.}~\bibnamefont
  {Champel}}, \bibinfo {author} {\bibfnamefont {J.~C.}\ \bibnamefont {Cuevas}},
  \bibinfo {author} {\bibfnamefont {J.}~\bibnamefont {Kopu}}, \ and\ \bibinfo
  {author} {\bibfnamefont {G.}~\bibnamefont {Sch{\"o}n}},\ }\href {\doibase
  10.1007/s10909-007-9329-6} {\bibfield  {journal} {\bibinfo  {journal} {J. Low
  Temp. Phys.}\ }\textbf {\bibinfo {volume} {147}},\ \bibinfo {pages} {457}
  (\bibinfo {year} {2007})}\BibitemShut {NoStop}%
\bibitem [{\citenamefont {Hara}\ and\ \citenamefont
  {Nagai}(1986)}]{hara_PTP_1986}%
  \BibitemOpen
  \bibfield  {author} {\bibinfo {author} {\bibfnamefont {J.}~\bibnamefont
  {Hara}}\ and\ \bibinfo {author} {\bibfnamefont {K.}~\bibnamefont {Nagai}},\
  }\href {\doibase 10.1143/PTP.76.1237} {\bibfield  {journal} {\bibinfo
  {journal} {Prog. Theor. Phys.}\ }\textbf {\bibinfo {volume} {76}},\ \bibinfo
  {pages} {1237} (\bibinfo {year} {1986})}\BibitemShut {NoStop}%
\bibitem [{Note1()}]{Note1}%
  \BibitemOpen
  \bibinfo {note} {The vortex shadow effect\cite { graser_PRL_2004,
  yokoyama_PRL_2008, silaev_PRB_2009} was discussed. However, odd-$\omega $
  pairs are not in the same location.}\BibitemShut {Stop}%
\bibitem [{\citenamefont {Ran}\ \emph {et~al.}(2019)\citenamefont {Ran},
  \citenamefont {Eckberg}, \citenamefont {Ding}, \citenamefont {Furukawa},
  \citenamefont {Metz}, \citenamefont {Saha}, \citenamefont {Liu},
  \citenamefont {Zic}, \citenamefont {Kim}, \citenamefont {Paglione},\ and\
  \citenamefont {Butch}}]{ran_science_2019}%
  \BibitemOpen
  \bibfield  {author} {\bibinfo {author} {\bibfnamefont {S.}~\bibnamefont
  {Ran}}, \bibinfo {author} {\bibfnamefont {C.}~\bibnamefont {Eckberg}},
  \bibinfo {author} {\bibfnamefont {Q.-P.}\ \bibnamefont {Ding}}, \bibinfo
  {author} {\bibfnamefont {Y.}~\bibnamefont {Furukawa}}, \bibinfo {author}
  {\bibfnamefont {T.}~\bibnamefont {Metz}}, \bibinfo {author} {\bibfnamefont
  {S.~R.}\ \bibnamefont {Saha}}, \bibinfo {author} {\bibfnamefont {I.-L.}\
  \bibnamefont {Liu}}, \bibinfo {author} {\bibfnamefont {M.}~\bibnamefont
  {Zic}}, \bibinfo {author} {\bibfnamefont {H.}~\bibnamefont {Kim}}, \bibinfo
  {author} {\bibfnamefont {J.}~\bibnamefont {Paglione}}, \ and\ \bibinfo
  {author} {\bibfnamefont {N.~P.}\ \bibnamefont {Butch}},\ }\href {\doibase
  10.1126/science.aav8645} {\bibfield  {journal} {\bibinfo  {journal}
  {Science}\ }\textbf {\bibinfo {volume} {365}},\ \bibinfo {pages} {684}
  (\bibinfo {year} {2019})}\BibitemShut {NoStop}%
\bibitem [{\citenamefont {Pustogow}\ \emph {et~al.}(2019)\citenamefont
  {Pustogow}, \citenamefont {Luo}, \citenamefont {Chronister}, \citenamefont
  {Su}, \citenamefont {Sokolov}, \citenamefont {Jerzembeck}, \citenamefont
  {Mackenzie}, \citenamefont {Hicks}, \citenamefont {Kikugawa}, \citenamefont
  {Raghu}, \citenamefont {Bauer},\ and\ \citenamefont
  {Brown}}]{pustogow_Nature_2019}%
  \BibitemOpen
  \bibfield  {author} {\bibinfo {author} {\bibfnamefont {A.}~\bibnamefont
  {Pustogow}}, \bibinfo {author} {\bibfnamefont {Y.}~\bibnamefont {Luo}},
  \bibinfo {author} {\bibfnamefont {A.}~\bibnamefont {Chronister}}, \bibinfo
  {author} {\bibfnamefont {Y.-S.}\ \bibnamefont {Su}}, \bibinfo {author}
  {\bibfnamefont {D.~A.}\ \bibnamefont {Sokolov}}, \bibinfo {author}
  {\bibfnamefont {F.}~\bibnamefont {Jerzembeck}}, \bibinfo {author}
  {\bibfnamefont {A.~P.}\ \bibnamefont {Mackenzie}}, \bibinfo {author}
  {\bibfnamefont {C.~W.}\ \bibnamefont {Hicks}}, \bibinfo {author}
  {\bibfnamefont {N.}~\bibnamefont {Kikugawa}}, \bibinfo {author}
  {\bibfnamefont {S.}~\bibnamefont {Raghu}}, \bibinfo {author} {\bibfnamefont
  {E.~D.}\ \bibnamefont {Bauer}}, \ and\ \bibinfo {author} {\bibfnamefont
  {S.~E.}\ \bibnamefont {Brown}},\ }\href {\doibase 10.1038/s41586-019-1596-2}
  {\bibfield  {journal} {\bibinfo  {journal} {Nature}\ }\textbf {\bibinfo
  {volume} {574}},\ \bibinfo {pages} {72} (\bibinfo {year} {2019})}\BibitemShut
  {NoStop}%
\bibitem [{\citenamefont {Suh}\ \emph {et~al.}(2020)\citenamefont {Suh},
  \citenamefont {Menke}, \citenamefont {Brydon}, \citenamefont {Timm},
  \citenamefont {Ramires},\ and\ \citenamefont
  {Agterberg}}]{Agterberg_PRR_2020}%
  \BibitemOpen
  \bibfield  {author} {\bibinfo {author} {\bibfnamefont {H.~G.}\ \bibnamefont
  {Suh}}, \bibinfo {author} {\bibfnamefont {H.}~\bibnamefont {Menke}}, \bibinfo
  {author} {\bibfnamefont {P.~M.~R.}\ \bibnamefont {Brydon}}, \bibinfo {author}
  {\bibfnamefont {C.}~\bibnamefont {Timm}}, \bibinfo {author} {\bibfnamefont
  {A.}~\bibnamefont {Ramires}}, \ and\ \bibinfo {author} {\bibfnamefont
  {D.~F.}\ \bibnamefont {Agterberg}},\ }\href {\doibase
  10.1103/PhysRevResearch.2.032023} {\bibfield  {journal} {\bibinfo  {journal}
  {Phys. Rev. Research}\ }\textbf {\bibinfo {volume} {2}},\ \bibinfo {pages}
  {032023} (\bibinfo {year} {2020})}\BibitemShut {NoStop}%
\bibitem [{\citenamefont {Grinenko}\ \emph
  {et~al.}(2021{\natexlab{a}})\citenamefont {Grinenko}, \citenamefont {Ghosh},
  \citenamefont {Sarkar}, \citenamefont {Orain}, \citenamefont {Nikitin},
  \citenamefont {Elender}, \citenamefont {Das}, \citenamefont {Guguchia},
  \citenamefont {Br\"{u}ckner}, \citenamefont {Barber}, \citenamefont {Park},
  \citenamefont {Kikugawa}, \citenamefont {Sokolov}, \citenamefont {Bobowski},
  \citenamefont {Miyoshi}, \citenamefont {Maeno}, \citenamefont {Mackenzie},
  \citenamefont {Luetkens}, \citenamefont {Hicks},\ and\ \citenamefont
  {Klauss}}]{Grinenko_NatPhys_2021}%
  \BibitemOpen
  \bibfield  {author} {\bibinfo {author} {\bibfnamefont {V.}~\bibnamefont
  {Grinenko}}, \bibinfo {author} {\bibfnamefont {S.}~\bibnamefont {Ghosh}},
  \bibinfo {author} {\bibfnamefont {R.}~\bibnamefont {Sarkar}}, \bibinfo
  {author} {\bibfnamefont {J.-C.}\ \bibnamefont {Orain}}, \bibinfo {author}
  {\bibfnamefont {A.}~\bibnamefont {Nikitin}}, \bibinfo {author} {\bibfnamefont
  {M.}~\bibnamefont {Elender}}, \bibinfo {author} {\bibfnamefont
  {D.}~\bibnamefont {Das}}, \bibinfo {author} {\bibfnamefont {Z.}~\bibnamefont
  {Guguchia}}, \bibinfo {author} {\bibfnamefont {F.}~\bibnamefont
  {Br\"{u}ckner}}, \bibinfo {author} {\bibfnamefont {M.~E.}\ \bibnamefont
  {Barber}}, \bibinfo {author} {\bibfnamefont {J.}~\bibnamefont {Park}},
  \bibinfo {author} {\bibfnamefont {N.}~\bibnamefont {Kikugawa}}, \bibinfo
  {author} {\bibfnamefont {D.~A.}\ \bibnamefont {Sokolov}}, \bibinfo {author}
  {\bibfnamefont {J.~S.}\ \bibnamefont {Bobowski}}, \bibinfo {author}
  {\bibfnamefont {T.}~\bibnamefont {Miyoshi}}, \bibinfo {author} {\bibfnamefont
  {Y.}~\bibnamefont {Maeno}}, \bibinfo {author} {\bibfnamefont {A.~P.}\
  \bibnamefont {Mackenzie}}, \bibinfo {author} {\bibfnamefont {H.}~\bibnamefont
  {Luetkens}}, \bibinfo {author} {\bibfnamefont {C.~W.}\ \bibnamefont {Hicks}},
  \ and\ \bibinfo {author} {\bibfnamefont {H.-H.}\ \bibnamefont {Klauss}},\
  }\href {\doibase https://doi.org/10.1038/s41567-021-01182-7} {\bibfield
  {journal} {\bibinfo  {journal} {Nat. Phys.}\ }\textbf {\bibinfo {volume}
  {17}},\ \bibinfo {pages} {748} (\bibinfo {year}
  {2021}{\natexlab{a}})}\BibitemShut {NoStop}%
\bibitem [{\citenamefont {Grinenko}\ \emph
  {et~al.}(2021{\natexlab{b}})\citenamefont {Grinenko}, \citenamefont {Das},
  \citenamefont {Gupta}, \citenamefont {Zinkl}, \citenamefont {Kikugawa},
  \citenamefont {Maeno}, \citenamefont {Hicks}, \citenamefont {Klauss},
  \citenamefont {Sigrist},\ and\ \citenamefont
  {Khasanov}}]{grinenko2021unsplit}%
  \BibitemOpen
  \bibfield  {author} {\bibinfo {author} {\bibfnamefont {V.}~\bibnamefont
  {Grinenko}}, \bibinfo {author} {\bibfnamefont {D.}~\bibnamefont {Das}},
  \bibinfo {author} {\bibfnamefont {R.}~\bibnamefont {Gupta}}, \bibinfo
  {author} {\bibfnamefont {B.}~\bibnamefont {Zinkl}}, \bibinfo {author}
  {\bibfnamefont {N.}~\bibnamefont {Kikugawa}}, \bibinfo {author}
  {\bibfnamefont {Y.}~\bibnamefont {Maeno}}, \bibinfo {author} {\bibfnamefont
  {C.~W.}\ \bibnamefont {Hicks}}, \bibinfo {author} {\bibfnamefont {H.-H.}\
  \bibnamefont {Klauss}}, \bibinfo {author} {\bibfnamefont {M.}~\bibnamefont
  {Sigrist}}, \ and\ \bibinfo {author} {\bibfnamefont {R.}~\bibnamefont
  {Khasanov}},\ }\href {https://www.nature.com/articles/s41467-021-24176-8}
  {\bibfield  {journal} {\bibinfo  {journal} {Nat. Commun.}\ }\textbf {\bibinfo
  {volume} {12}},\ \bibinfo {pages} {1} (\bibinfo {year}
  {2021}{\natexlab{b}})}\BibitemShut {NoStop}%
\bibitem [{\citenamefont {Kobayashi}\ \emph {et~al.}(2015)\citenamefont
  {Kobayashi}, \citenamefont {Tanaka},\ and\ \citenamefont
  {Sato}}]{Shingo_PRB_2015}%
  \BibitemOpen
  \bibfield  {author} {\bibinfo {author} {\bibfnamefont {S.}~\bibnamefont
  {Kobayashi}}, \bibinfo {author} {\bibfnamefont {Y.}~\bibnamefont {Tanaka}}, \
  and\ \bibinfo {author} {\bibfnamefont {M.}~\bibnamefont {Sato}},\ }\href
  {\doibase 10.1103/PhysRevB.92.214514} {\bibfield  {journal} {\bibinfo
  {journal} {Phys. Rev. B}\ }\textbf {\bibinfo {volume} {92}},\ \bibinfo
  {pages} {214514} (\bibinfo {year} {2015})}\BibitemShut {NoStop}%
\bibitem [{\citenamefont {Tamura}\ \emph {et~al.}(2017)\citenamefont {Tamura},
  \citenamefont {Kobayashi}, \citenamefont {Bo},\ and\ \citenamefont
  {Tanaka}}]{TamuShun_PRB_2017}%
  \BibitemOpen
  \bibfield  {author} {\bibinfo {author} {\bibfnamefont {S.}~\bibnamefont
  {Tamura}}, \bibinfo {author} {\bibfnamefont {S.}~\bibnamefont {Kobayashi}},
  \bibinfo {author} {\bibfnamefont {L.}~\bibnamefont {Bo}}, \ and\ \bibinfo
  {author} {\bibfnamefont {Y.}~\bibnamefont {Tanaka}},\ }\href {\doibase
  10.1103/PhysRevB.95.104511} {\bibfield  {journal} {\bibinfo  {journal} {Phys.
  Rev. B}\ }\textbf {\bibinfo {volume} {95}},\ \bibinfo {pages} {104511}
  (\bibinfo {year} {2017})}\BibitemShut {NoStop}%
\bibitem [{\citenamefont {Suzuki}\ \emph {et~al.}(2020)\citenamefont {Suzuki},
  \citenamefont {Sato},\ and\ \citenamefont {Tanaka}}]{suzuki_PRB_2020}%
  \BibitemOpen
  \bibfield  {author} {\bibinfo {author} {\bibfnamefont {S.-I.}\ \bibnamefont
  {Suzuki}}, \bibinfo {author} {\bibfnamefont {M.}~\bibnamefont {Sato}}, \ and\
  \bibinfo {author} {\bibfnamefont {Y.}~\bibnamefont {Tanaka}},\ }\href
  {\doibase 10.1103/PhysRevB.101.054505} {\bibfield  {journal} {\bibinfo
  {journal} {Phys. Rev. B}\ }\textbf {\bibinfo {volume} {101}},\ \bibinfo
  {pages} {054505} (\bibinfo {year} {2020})}\BibitemShut {NoStop}%
\bibitem [{\citenamefont {Suzuki}\ \emph {et~al.}(2022)\citenamefont {Suzuki},
  \citenamefont {Ikegaya},\ and\ \citenamefont {Golubov}}]{suzuki2022}%
  \BibitemOpen
  \bibfield  {author} {\bibinfo {author} {\bibfnamefont {S.-I.}\ \bibnamefont
  {Suzuki}}, \bibinfo {author} {\bibfnamefont {S.}~\bibnamefont {Ikegaya}}, \
  and\ \bibinfo {author} {\bibfnamefont {A.~A.}\ \bibnamefont {Golubov}},\
  }\href {https://arxiv.org/abs/2207.04094} {\bibfield  {journal} {\bibinfo
  {journal} {arXiv:2207.04094}\ } (\bibinfo {year} {2022})}\BibitemShut
  {NoStop}%
\bibitem [{\citenamefont {Hess}\ \emph {et~al.}(1989)\citenamefont {Hess},
  \citenamefont {Robinson}, \citenamefont {Dynes}, \citenamefont {Valles},\
  and\ \citenamefont {Waszczak}}]{hess_PRL_1989}%
  \BibitemOpen
  \bibfield  {author} {\bibinfo {author} {\bibfnamefont {H.~F.}\ \bibnamefont
  {Hess}}, \bibinfo {author} {\bibfnamefont {R.~B.}\ \bibnamefont {Robinson}},
  \bibinfo {author} {\bibfnamefont {R.~C.}\ \bibnamefont {Dynes}}, \bibinfo
  {author} {\bibfnamefont {J.~M.}\ \bibnamefont {Valles}}, \ and\ \bibinfo
  {author} {\bibfnamefont {J.~V.}\ \bibnamefont {Waszczak}},\ }\href {\doibase
  10.1103/PhysRevLett.62.214} {\bibfield  {journal} {\bibinfo  {journal} {Phys.
  Rev. Lett.}\ }\textbf {\bibinfo {volume} {62}},\ \bibinfo {pages} {214}
  (\bibinfo {year} {1989})}\BibitemShut {NoStop}%
\bibitem [{\citenamefont {Kirtley}\ and\ \citenamefont
  {Wikswo~Jr}(1999)}]{squid1}%
  \BibitemOpen
  \bibfield  {author} {\bibinfo {author} {\bibfnamefont {J.~R.}\ \bibnamefont
  {Kirtley}}\ and\ \bibinfo {author} {\bibfnamefont {J.~P.}\ \bibnamefont
  {Wikswo~Jr}},\ }\href@noop {} {\bibfield  {journal} {\bibinfo  {journal}
  {Annual Review of Materials Science}\ }\textbf {\bibinfo {volume} {29}},\
  \bibinfo {pages} {117} (\bibinfo {year} {1999})}\BibitemShut {NoStop}%
\bibitem [{\citenamefont {Wells}\ \emph {et~al.}(2015)\citenamefont {Wells},
  \citenamefont {Pan}, \citenamefont {Wang}, \citenamefont {Fedoseev},\ and\
  \citenamefont {Hilgenkamp}}]{squid2}%
  \BibitemOpen
  \bibfield  {author} {\bibinfo {author} {\bibfnamefont {F.~S.}\ \bibnamefont
  {Wells}}, \bibinfo {author} {\bibfnamefont {A.~V.}\ \bibnamefont {Pan}},
  \bibinfo {author} {\bibfnamefont {X.~R.}\ \bibnamefont {Wang}}, \bibinfo
  {author} {\bibfnamefont {S.~A.}\ \bibnamefont {Fedoseev}}, \ and\ \bibinfo
  {author} {\bibfnamefont {H.}~\bibnamefont {Hilgenkamp}},\ }\href@noop {}
  {\bibfield  {journal} {\bibinfo  {journal} {Scientific reports}\ }\textbf
  {\bibinfo {volume} {5}},\ \bibinfo {pages} {1} (\bibinfo {year}
  {2015})}\BibitemShut {NoStop}%
\bibitem [{\citenamefont {Caroli}\ \emph {et~al.}(1964)\citenamefont {Caroli},
  \citenamefont {{De Gennes}},\ and\ \citenamefont
  {Matricon}}]{caroli_PL_1964}%
  \BibitemOpen
  \bibfield  {author} {\bibinfo {author} {\bibfnamefont {C.}~\bibnamefont
  {Caroli}}, \bibinfo {author} {\bibfnamefont {P.}~\bibnamefont {{De Gennes}}},
  \ and\ \bibinfo {author} {\bibfnamefont {J.}~\bibnamefont {Matricon}},\
  }\href {\doibase https://doi.org/10.1016/0031-9163(64)90375-0} {\bibfield
  {journal} {\bibinfo  {journal} {Phys. Lett.}\ }\textbf {\bibinfo {volume}
  {9}},\ \bibinfo {pages} {307} (\bibinfo {year} {1964})}\BibitemShut {NoStop}%
\bibitem [{\citenamefont {Gygi}\ and\ \citenamefont
  {Schl\"uter}(1991)}]{gygi_PRB_1991}%
  \BibitemOpen
  \bibfield  {author} {\bibinfo {author} {\bibfnamefont {F.}~\bibnamefont
  {Gygi}}\ and\ \bibinfo {author} {\bibfnamefont {M.}~\bibnamefont
  {Schl\"uter}},\ }\href {\doibase 10.1103/PhysRevB.43.7609} {\bibfield
  {journal} {\bibinfo  {journal} {Phys. Rev. B}\ }\textbf {\bibinfo {volume}
  {43}},\ \bibinfo {pages} {7609} (\bibinfo {year} {1991})}\BibitemShut
  {NoStop}%
\bibitem [{\citenamefont {Kramer}\ and\ \citenamefont
  {Pesch}(1974)}]{kramer_ZP_1974}%
  \BibitemOpen
  \bibfield  {author} {\bibinfo {author} {\bibfnamefont {L.}~\bibnamefont
  {Kramer}}\ and\ \bibinfo {author} {\bibfnamefont {W.}~\bibnamefont {Pesch}},\
  }\href {\doibase 10.1007/BF01668869} {\bibfield  {journal} {\bibinfo
  {journal} {Z. Physik}\ }\textbf {\bibinfo {volume} {269}},\ \bibinfo {pages}
  {59} (\bibinfo {year} {1974})}\BibitemShut {NoStop}%
\bibitem [{\citenamefont {Nagai}\ \emph {et~al.}(2006)\citenamefont {Nagai},
  \citenamefont {Ueno}, \citenamefont {Kato},\ and\ \citenamefont
  {Hayashi}}]{nagai_JPSJ_2006}%
  \BibitemOpen
  \bibfield  {author} {\bibinfo {author} {\bibfnamefont {Y.}~\bibnamefont
  {Nagai}}, \bibinfo {author} {\bibfnamefont {Y.}~\bibnamefont {Ueno}},
  \bibinfo {author} {\bibfnamefont {Y.}~\bibnamefont {Kato}}, \ and\ \bibinfo
  {author} {\bibfnamefont {N.}~\bibnamefont {Hayashi}},\ }\href {\doibase
  10.1143/jpsj.75.104701} {\bibfield  {journal} {\bibinfo  {journal} {J. Phys.
  Soc. Jpn.}\ }\textbf {\bibinfo {volume} {75}},\ \bibinfo {pages} {104701}
  (\bibinfo {year} {2006})}\BibitemShut {NoStop}%
\bibitem [{\citenamefont {Nagai}\ and\ \citenamefont
  {Kato}(2019)}]{nagai_JPSJ_2019}%
  \BibitemOpen
  \bibfield  {author} {\bibinfo {author} {\bibfnamefont {Y.}~\bibnamefont
  {Nagai}}\ and\ \bibinfo {author} {\bibfnamefont {Y.}~\bibnamefont {Kato}},\
  }\href {\doibase 10.7566/JPSJ.88.054707} {\bibfield  {journal} {\bibinfo
  {journal} {J. Phys. Soc. Jpn.}\ }\textbf {\bibinfo {volume} {88}},\ \bibinfo
  {pages} {054707} (\bibinfo {year} {2019})}\BibitemShut {NoStop}%
\bibitem [{\citenamefont {Eilenberger}(1968)}]{eilenberger_ZP_1968}%
  \BibitemOpen
  \bibfield  {author} {\bibinfo {author} {\bibfnamefont {G.}~\bibnamefont
  {Eilenberger}},\ }\href {\doibase 10.1007/BF01379803} {\bibfield  {journal}
  {\bibinfo  {journal} {Z. Physik}\ }\textbf {\bibinfo {volume} {214}},\
  \bibinfo {pages} {195} (\bibinfo {year} {1968})}\BibitemShut {NoStop}%
\bibitem [{\citenamefont {Eilenberger}\ and\ \citenamefont
  {B{\"u}ttner}(1969)}]{eilenberger_ZP_1969}%
  \BibitemOpen
  \bibfield  {author} {\bibinfo {author} {\bibfnamefont {G.}~\bibnamefont
  {Eilenberger}}\ and\ \bibinfo {author} {\bibfnamefont {H.}~\bibnamefont
  {B{\"u}ttner}},\ }\href {\doibase 10.1007/BF01393061} {\bibfield  {journal}
  {\bibinfo  {journal} {Z. Physik}\ }\textbf {\bibinfo {volume} {224}},\
  \bibinfo {pages} {335} (\bibinfo {year} {1969})}\BibitemShut {NoStop}%
\bibitem [{\citenamefont {Larkin}\ and\ \citenamefont
  {Ovchinnikov}(1969)}]{larkin1969quasiclassical}%
  \BibitemOpen
  \bibfield  {author} {\bibinfo {author} {\bibfnamefont {A.}~\bibnamefont
  {Larkin}}\ and\ \bibinfo {author} {\bibfnamefont {Y.~N.}\ \bibnamefont
  {Ovchinnikov}},\ }\href@noop {} {\bibfield  {journal} {\bibinfo  {journal}
  {J. Exp. Theor. Phys.}\ }\textbf {\bibinfo {volume} {28}},\ \bibinfo {pages}
  {1200} (\bibinfo {year} {1969})}\BibitemShut {NoStop}%
\bibitem [{\citenamefont {Schopohl}\ and\ \citenamefont
  {Maki}(1995)}]{schopohl_PRB_1995}%
  \BibitemOpen
  \bibfield  {author} {\bibinfo {author} {\bibfnamefont {N.}~\bibnamefont
  {Schopohl}}\ and\ \bibinfo {author} {\bibfnamefont {K.}~\bibnamefont
  {Maki}},\ }\href {\doibase 10.1103/PhysRevB.52.490} {\bibfield  {journal}
  {\bibinfo  {journal} {Phys. Rev. B}\ }\textbf {\bibinfo {volume} {52}},\
  \bibinfo {pages} {490} (\bibinfo {year} {1995})}\BibitemShut {NoStop}%
\bibitem [{\citenamefont {Schopohl}(1998)}]{schopohl1998transformation}%
  \BibitemOpen
  \bibfield  {author} {\bibinfo {author} {\bibfnamefont {N.}~\bibnamefont
  {Schopohl}},\ }\href@noop {} {\bibfield  {journal} {\bibinfo  {journal}
  {arXiv preprint cond-mat/9804064}\ } (\bibinfo {year} {1998})}\BibitemShut
  {NoStop}%
\bibitem [{\citenamefont
  {Higashitani}(2014{\natexlab{b}})}]{Higashitani_JPSJ_2014}%
  \BibitemOpen
  \bibfield  {author} {\bibinfo {author} {\bibfnamefont {S.}~\bibnamefont
  {Higashitani}},\ }\href {\doibase 10.7566/JPSJ.83.075002} {\bibfield
  {journal} {\bibinfo  {journal} {J. Phys. Soc. Jpn.}\ }\textbf {\bibinfo
  {volume} {83}},\ \bibinfo {pages} {075002} (\bibinfo {year}
  {2014}{\natexlab{b}})}\BibitemShut {NoStop}%
\bibitem [{Note2()}]{Note2}%
  \BibitemOpen
  \bibinfo {note} {At a surface of an SC, the incoming and outgoing
  quasiparticles interfere. These two quasiparticles feel, in general,
  different pair potentials depending on the momentum: $\Delta (k_z,\protect
  \boldsymbol {k}_\parallel )$ and $\Delta (-k_z,\protect \boldsymbol
  {k}_\parallel )$ where the surface is perpendicular to the $z$ axis. In the
  $p_z$-wave SC, the phase difference between these two pair potentials is $\pi
  $: $\Delta (k_z,\protect \boldsymbol {k}_\parallel ) = -\Delta (-k_z,\protect
  \boldsymbol {k}_\parallel )$. The $\pi $-phase shift results in forming the
  surface bound state at the zero energy.}\BibitemShut {Stop}%
\bibitem [{\citenamefont {Hu}(1994)}]{hu_PRL_1994}%
  \BibitemOpen
  \bibfield  {author} {\bibinfo {author} {\bibfnamefont {C.-R.}\ \bibnamefont
  {Hu}},\ }\href {\doibase 10.1103/PhysRevLett.72.1526} {\bibfield  {journal}
  {\bibinfo  {journal} {Phys. Rev. Lett.}\ }\textbf {\bibinfo {volume} {72}},\
  \bibinfo {pages} {1526} (\bibinfo {year} {1994})}\BibitemShut {NoStop}%
\bibitem [{\citenamefont {Kashiwaya}\ and\ \citenamefont
  {Tanaka}(2000)}]{kashiwaya_RPP_2000}%
  \BibitemOpen
  \bibfield  {author} {\bibinfo {author} {\bibfnamefont {S.}~\bibnamefont
  {Kashiwaya}}\ and\ \bibinfo {author} {\bibfnamefont {Y.}~\bibnamefont
  {Tanaka}},\ }\href {\doibase 10.1088/0034-4885/63/10/202} {\bibfield
  {journal} {\bibinfo  {journal} {Rep. Prog. Phys.}\ }\textbf {\bibinfo
  {volume} {63}},\ \bibinfo {pages} {1641} (\bibinfo {year}
  {2000})}\BibitemShut {NoStop}%
\bibitem [{\citenamefont {Sato}\ \emph {et~al.}(2011)\citenamefont {Sato},
  \citenamefont {Tanaka}, \citenamefont {Yada},\ and\ \citenamefont
  {Yokoyama}}]{sato_PRB_2011}%
  \BibitemOpen
  \bibfield  {author} {\bibinfo {author} {\bibfnamefont {M.}~\bibnamefont
  {Sato}}, \bibinfo {author} {\bibfnamefont {Y.}~\bibnamefont {Tanaka}},
  \bibinfo {author} {\bibfnamefont {K.}~\bibnamefont {Yada}}, \ and\ \bibinfo
  {author} {\bibfnamefont {T.}~\bibnamefont {Yokoyama}},\ }\href {\doibase
  10.1103/PhysRevB.83.224511} {\bibfield  {journal} {\bibinfo  {journal} {Phys.
  Rev. B}\ }\textbf {\bibinfo {volume} {83}},\ \bibinfo {pages} {224511}
  (\bibinfo {year} {2011})}\BibitemShut {NoStop}%
\bibitem [{Note3()}]{Note3}%
  \BibitemOpen
  \bibinfo {note} {The energy levels for the CdGM modes are given by $E=\pm
  (n-1/2) \Delta _\infty ^2/E_F$. In the quasiclassical regime, the lowest CdGM
  mode seem to sit at $E=0$ because $\Delta /E_F \ll 1$ is assumed in the
  quasiclassical theory.}\BibitemShut {Stop}%
\bibitem [{\citenamefont {Graser}\ \emph {et~al.}(2004)\citenamefont {Graser},
  \citenamefont {Iniotakis}, \citenamefont {Dahm},\ and\ \citenamefont
  {Schopohl}}]{graser_PRL_2004}%
  \BibitemOpen
  \bibfield  {author} {\bibinfo {author} {\bibfnamefont {S.}~\bibnamefont
  {Graser}}, \bibinfo {author} {\bibfnamefont {C.}~\bibnamefont {Iniotakis}},
  \bibinfo {author} {\bibfnamefont {T.}~\bibnamefont {Dahm}}, \ and\ \bibinfo
  {author} {\bibfnamefont {N.}~\bibnamefont {Schopohl}},\ }\href {\doibase
  10.1103/PhysRevLett.93.247001} {\bibfield  {journal} {\bibinfo  {journal}
  {Phys. Rev. Lett.}\ }\textbf {\bibinfo {volume} {93}},\ \bibinfo {pages}
  {247001} (\bibinfo {year} {2004})}\BibitemShut {NoStop}%
\bibitem [{\citenamefont {Yokoyama}\ \emph
  {et~al.}(2008{\natexlab{b}})\citenamefont {Yokoyama}, \citenamefont
  {Iniotakis}, \citenamefont {Tanaka},\ and\ \citenamefont
  {Sigrist}}]{yokoyama_PRL_2008}%
  \BibitemOpen
  \bibfield  {author} {\bibinfo {author} {\bibfnamefont {T.}~\bibnamefont
  {Yokoyama}}, \bibinfo {author} {\bibfnamefont {C.}~\bibnamefont {Iniotakis}},
  \bibinfo {author} {\bibfnamefont {Y.}~\bibnamefont {Tanaka}}, \ and\ \bibinfo
  {author} {\bibfnamefont {M.}~\bibnamefont {Sigrist}},\ }\href {\doibase
  10.1103/PhysRevLett.100.177002} {\bibfield  {journal} {\bibinfo  {journal}
  {Phys. Rev. Lett.}\ }\textbf {\bibinfo {volume} {100}},\ \bibinfo {pages}
  {177002} (\bibinfo {year} {2008}{\natexlab{b}})}\BibitemShut {NoStop}%
\bibitem [{\citenamefont {Silaev}\ \emph {et~al.}(2009)\citenamefont {Silaev},
  \citenamefont {Yokoyama}, \citenamefont {Linder}, \citenamefont {Tanaka},\
  and\ \citenamefont {Sudb\o{}}}]{silaev_PRB_2009}%
  \BibitemOpen
  \bibfield  {author} {\bibinfo {author} {\bibfnamefont {M.~A.}\ \bibnamefont
  {Silaev}}, \bibinfo {author} {\bibfnamefont {T.}~\bibnamefont {Yokoyama}},
  \bibinfo {author} {\bibfnamefont {J.}~\bibnamefont {Linder}}, \bibinfo
  {author} {\bibfnamefont {Y.}~\bibnamefont {Tanaka}}, \ and\ \bibinfo {author}
  {\bibfnamefont {A.}~\bibnamefont {Sudb\o{}}},\ }\href {\doibase
  10.1103/PhysRevB.79.054508} {\bibfield  {journal} {\bibinfo  {journal} {Phys.
  Rev. B}\ }\textbf {\bibinfo {volume} {79}},\ \bibinfo {pages} {054508}
  (\bibinfo {year} {2009})}\BibitemShut {NoStop}%
\end{thebibliography}%

\end{document}